\documentclass[preprint,12pt,3p]{elsarticle}
\usepackage{graphicx}
\usepackage{amssymb}
\usepackage{caption}
\usepackage{subcaption}
\usepackage{url}
\usepackage{amsmath}
\graphicspath{{./images_revised3/}}

\usepackage{tikz}
\tikzset{
  treenode/.style = {shape=rectangle, rounded corners,
                     draw, align=center,
                     top color=white, bottom color=blue!20},
  root/.style     = {treenode, font=\Large, bottom color=red!30},
  env/.style      = {treenode, font=\ttfamily\normalsize},
  dummy/.style    = {circle,draw}
}

\journal{AMS: Women in Mathematics}

\begin{document}

\begin{frontmatter}

\title{On the dynamic suction pumping of blood cells in tubular hearts}
\tnotetext[label0]{This is only an example}

\author[label1]{Nicholas A. Battista\corref{cor1}\fnref{label3}}
\address[label1]{Department of Mathematics, CB 3250, University of North Carolina, Chapel Hill, NC, 27599}

\cortext[cor1]{I am corresponding author}

\ead{nick.battista@unc.edu}
\ead[url]{battista.web.unc.edu}

\author[label1,label2]{Andrea N. Lane}
\address[label2]{Department of Biostatistics, UNC Gillings School of Global Public Health, Chapel Hill, NC, 27599}
\ead{anlane@live.unc.edu}


\author[label1,label3]{Laura A. Miller}
\address[label3]{Department of Biology, 3280, University of North Carolina, Chapel Hill, NC, 27599}
\ead{lam9@unc.edu}

\begin{abstract}
Around the third week after gestation in embryonic development, the human heart consists only of a valvless tube, unlike a fully developed adult heart, which is multi-chambered. At this stage in development, the heart valves have not formed and so net flow of blood through the heart must be driven by a different mechanism. It is hypothesized that there are two possible mechanisms that drive blood flow at this stage - Liebau pumping (dynamic suction pumping or valveless pumping) and peristaltic pumping. We implement the immersed boundary method with adaptive mesh refinement (IBAMR) to numerically study the effect of hematocrit on the circulation around a valveless. Both peristalsis and dynamic suction pumping are considered. In the case of dynamic suction pumping, the heart and circulatory system is simplified as a flexible tube attached to a relatively rigid racetrack. For some Womersley number ($Wo$) regimes, there is significant net flow around the racetrack. We find that the addition of flexible blood cells does not significantly affect flow rates within the tube for $Wo\leq10$. On the other hand, peristalsis consistently drives blood around the racetrack for all $Wo$ and for all hematocrit considered.
\end{abstract}



\end{frontmatter}



%

%
%

\section{Introduction}

The Liebau pump (dynamic suction pump), first described in 1954 \cite{Meier:2011}, was studied as a novel way to pump water. It has not been until the past 20 years that scientists started looking at the pump as a valveless pumping mechanism in many biological systems and biomedical applications, including microelectromechanical systems (MEMs) and micro-fluidic devices. Direct applications of such pumps include tissue engineering, implantable micro electrodes, and drug delivery \cite{Lee:2004,Chang:2007,Lee:2008,Meier:2011}. 

\begin{figure}
    \centering
    \includegraphics[scale=2.1]{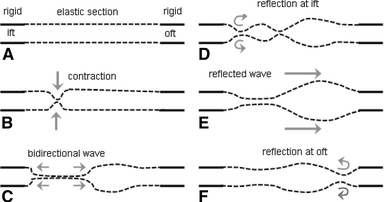}
    \caption{Schematic diagram illustrating dynamic suction pumping \cite{Santhanakrishnan:2011}. (A) The flexible tube is at rest. (B) Active contraction of the tube in a non-central location along the tube. (C) Contraction induces an elastic passive bidirectional wave to propagate along the tube. (D) Wave reflects off rigid portion of the tube on side nearest to contraction point. (E) The reflected wave travels down the tube. (F) The waves reflect off the rigid section at the far side of the tube. Notice the the reflected wave amplitude is smaller than the reflected wave off the other end.}
    \label{fig:DSP_Schematic}
\end{figure}

With extensive industrial applications, dynamic suction pumping (DSP) has proven to be a suitable means of transport for fluids and other materials in a valveless system, for scales of $Wo>1$ \cite{Baird:2014}. DSP can be most simply described by an isolated region of actuation, located asymmetrically along a flexible tube with stiffer ends. Flexibility of the tube is required to allow passive elastic traveling waves, which augment bulk transport throughout the system. The rigid ends of the tube cause reflections of the elastic waves, which when coupled with an asymmetric actuation point, can promote unidirectional flow. DSP is illustrated in Figure \ref{fig:DSP_Schematic}.

Due to a coupling between the system's geometry, material properties of the tube wall, and pumping mechanics, there is a complex, nonlinear relationship between volumetric flow rate and pumping frequency \cite{Baird:2014,Bringley:2008,Hickerson:2005}. Analytic models of DSP have been developed to address this relationship \cite{Ottsen:2003,Auerbach:2004,Manopoulos:2006,Samson:2007,Bringley:2008,Babbs:2010}. Most models use simplifications such as the inviscid assumption, long wave approximation, small contraction amplitude, one-dimensional flow. Furthermore, no analytical model has described flow reversals, which can occur with changes in the pumping frequency. Relaxing many of these assumptions, physical experiments have been performed to better understand DSP \cite{Hickerson:2005,HickersonThesis:2005,Bringley:2008,Meier:2011}, as well as \emph{in silico} investigations \cite{Jung:1999,Jung:2001,Avrahami:2008,Baird:2014,BairdThesis:2014}.  Most of these experimental and computational studies focus on the `high' $Wo$ regime ($Wo>>1$).  
       
The vertebrate embryonic heart is a valveless tube, similar to those in various invertebrates, such as urochordates and cephalochordates \cite{Kriebel:1967,Randall:1980}. Historically, the pumping mechanism in these hearts has been described as peristalsis \cite{Kriebel:1967, Santhanakrishnan:2011}. More recently, DSP has been proposed as a novel cardiac pumping mechanism for the vertebrate embryonic heart by Kenner \emph{et. al.} in $2000$ \cite{Kenner:2000}, and was later declared the main pumping mechanism in vertebrate embryonic hearts by Fourhar \emph{et. al.} in $2004$ \cite{Forouhar:2006}. Debate over which is the actual pumping mechanism of the embryonic heart continues today, with the possibility that the mechanism may vary between species or may be some hybrid of both mechanisms \cite{Waldrop:15BMMB,Manner:2010}.  

\begin{figure}[!h]
  \centering
  \includegraphics[scale=2.275]{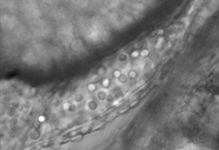}
  \caption{The embryonic heart tube of a Zebrafish 30 hpf courtesy of \cite{Maes:2011}. Spherical blood cells are seen within the tubular heart. The heart tube is roughly $5$ blood cells thick in diameter.}
  \label{Maes30hpf}
\end{figure}

Although the size of the blood cells during the tubular heart stage is on the same order of magnitude as the tube itself, previous work with numerical, analytical, and physical have not considered their presence. Given their size ($d \approx 4\ \mu m$) and volume fraction (hematocrit) that ranges from 0-40\%, it is likely that the blood cells are having some effect on the flow. When the first coordinated myocardial contractions begin to drive blood flow, the embryonic blood lacks blood cells.  However, as the heart tube stage progresses, the hematocrit (the volume fraction of blood cells) becomes present, as seen in Figure \ref{Maes30hpf}, and increases linearly during development \cite{Roubaie:2011}. Hematocrit may play a role in the distribution of forces along the endothelial lining that contribute to the shaping and growth of the heart. 

The purpose of this paper is explore the performance DSP and peristalsis when blood cells are added to the flow. In particular, a central goal is to quantify the relationship between the magnitude of flow and the hematocrit in tubular hearts over a range of Womersley Numbers, $Wo$. While the vertebrate tubular heart is on the order of tens of microns ($Wo<1$) \cite{Baird:2014}, the tubular hearts of many invertebrates span from the tens to hundreds of microns ($Wo<1$), e.g., sea squirts, to salps hearts on the order of millimeters ($Wo>1$) \cite{Kriebel:1967}. These ranges of $Wo$ naturally lend themselves to numerical study via the immersed boundary method. The particular geometry for  the computational models will be based upon experimental data from zebrafish, \emph{Danio rerio}, embryonic tubular hearts. 



%
%
\section{The Immersed Boundary Method}
The immersed boundary method (IBM) is a numerical method developed to solve viscous incompressible fluid dynamic problems with an immersed elastic structure \cite{Peskin:2002,Mittal:2005}. Since its development in the 1970s by Charles Peskin \cite{Peskin:1977}, it has been applied to a wide spectrum of biomathematical models, ranging from blood flow through the heart \cite{Peskin:1977,Peskin:2002}, aquatic locomotion \cite{Hieber:2008}, insect flight \cite{Miller:2004,Miller:2009,SJones:2015}, to plant biomechanics \cite{Miller:2012,Zhu:2011}. 

The power of this method is that it can be used to describe flow around complicated time-dependent geometries using a regular Cartesian discretization of the fluid domain. The elastic fibers describing the structure are discretized on a moving curvilinear mesh defined in the Lagrangian frame. The fluid and elastic fibers constitute a coupled system, in which the structure moves at the local fluid velocity and the structure applies a singular force of delta-layered thickness to the fluid.

We used an adaptive and parallelized version of the immersed boundary method, IBAMR \cite{BGriffithIBAMR,Griffith:2007}. IBAMR is a C++ framework that provides discretization and solver infrastructure for partial differential equations on block-structured locally refined Eulerian grids \cite{MJBerger84,MJBerger89} and on Lagrangian (structural) meshes. Adaptive mesh refinement (AMR) allows for better resolved dynamics between the fibers and the fluid by increasing grid resolution in areas of the domain that contain an immersed structure or where the vorticity exceeds some threshold. AMR also improves computational efficiency by decreasing grid resolution in areas of the domain that do not require a high level of resolution. IBAMR also includes infrastructure for coupling Eulerian and Lagrangian representations of the fluid and structure, respectively.  

The Eulerian grid on which the Navier-Stokes equations were solved was locally refined near the immersed boundaries and regions of vorticity with a threshold of $|\omega| > 0.20$. This Cartesian grid was organized as a hierarchy of four nested grid levels, and the finest grid was assigned a spatial step size of $dx = D/1024$, where $D$ is the length of the domain. The ratio of the spatial step size on each grid relative to the next coarsest grid was 1:4. The temporal resolution was varied to ensure stability. Each Lagrangian point of the immersed structure was chosen to be $\frac{D}{2048}$ apart, that is, twice the resolution of the finest fluid grid.

%
%

\subsection{Equations of the IBM}
Assume that the immsersed boundary is described on a curvilinear, Lagrangian mesh, $S$, that is free to move. The fluid is described on a fixed Cartesian, Eulerian grid, $\Omega$, that has periodic boundary conditions. Given the size of the domain and the localization of the flow to the tube, the boundary conditions do not significantly affect the fluid motion. The governing equations for the fluid, the Navier-Stokes equations, are given by
%
%
\begin{align}
\label{momentum} \rho \left[ \frac{\partial {\bf{u}} }{\partial t}( {\bf{x}},t ) +  \left({\bf{u}}( {\bf{x}},t )\cdot \nabla \right){\bf{u}}( {\bf{x}},t ) \right] &= -\nabla p( {\bf{x}},t ) + \mu \Delta{\bf{u}}( {\bf{x}},t ) + {\bf{f}}( {\bf{x}},t )  \\
\label{incompressible} \nabla&\cdot{\bf{u}}({\bf{x}},t) = 0.
\end{align}
Eqs.(\ref{momentum}) and (\ref{incompressible}) are the Navier-Stokes equations written in Eulerian form, where Eq.(\ref{momentum}) is the conservation of momentum for a fluid and Eq.(\ref{incompressible}) is the conservation of mass, i.e., incompressibility condition. The two constant parameters in these equations are the fluid density, $\rho$, and the dynamic viscosity of the fluid, $\mu$. The fluid velocity, ${\bf{u}}( {\bf{x}},t )$, pressure, $p( {\bf{x}},t )$, and body force, ${\bf{f}}( {\bf{x}},t)$, are unknown spatial time-dependent functions of the Eulerian coordinate, ${\bf{x}}$, and time, $t$. The body force describes the transfer of momentum onto the fluid due to the restoring forces arising from deformations of the elastic structure. It is this term, ${\bf{f}}( {\bf{x}},t)$, that is unique to the particular model being studied. 

The material properties of the structure may be modeled to resist to bending, stretching, and displacement from a tethered position. Other forces that can have been modeled include the action of virtual muscles, electrostatic (contact) forces, molecular bonds, and other external forces \cite{Peskin:2002,Tytell:2010,Mahur:2012,Fogelson:2008}. The immersed structure may deform due to bending forces and/or stretching and compression forces. In this paper, elastic  forces are calculated as beams that may undergoe large deformations and Hookean springs, i.e.,
%
%
\begin{eqnarray}
\label{beam_force} \mathbf{F}_{beam} &=& -k_{beam} \frac{\partial^4}{\partial s^4}\Big( \mathbf{X}(s,t) - \mathbf{X}_B(s) \Big) \\
\label{spring_force} \mathbf{F}_{spring} &=& - k_{spring} \left( 1 - \frac{R_L}{\left|\left| \mathbf{X}_{S} - \mathbf{X}_M \right|\right| } \right) \cdot \left( \mathbf{X}_M - \mathbf{X}_S \right),
\end{eqnarray}
Eq.(\ref{beam_force}) is the beam equation, which describes forces arising from bending of the elastic structure. Eq.(\ref{spring_force}) describes the force generated from stretching and compression of the structure. The parameters, $k_{beam}$ and $k_{spring}$, are the stiffness coefficients of the beam and spring, respectively, and $R_L$ is the resting length of the Hookean spring. The variables $\mathbf{X}_M$ and $\mathbf{X}_S$ give the positions in Cartesian coordinates of the master and slave nodes in the spring formulations, respectively, $\mathbf{X}_B(s)$ describes the deviation from the preferred curvature of the structure. In all simulations, $\mathbf{X}_B(s) = 0$ along the straight portion of the tube.

A target point formulation can be used to tether the structure or subset thereof in place, holding the Lagrangian mesh in a preferred position that may be time dependent. An immersed boundary point with position $\mathbf{X}(s,t)$ that is tethered to a target point, with position $\mathbf{Y}(s,t)$ undergoes a penalty force that is proportional to the displacement between them. The force that results is given by the equation for a linear spring with zero resting length, 
\begin{equation}
\label{target_force} \mathbf{F}_{target}= -\kappa_{target}\left( {\bf{X}}(s,t) - {\bf{Y}}(s,t) \right), \\
\end{equation}
where $k_{target}$ is the stiffness coefficient of the target point springs. $k_{target}$ can be varied to control the deviation allowed between the actual location of the boundary and its preferred position. The total deformation force that will be applied to the fluid is a sum of the above forces, 

\begin{equation}
\label{deform_force} {\bf{F}}(s,t)= \mathbf{F}_{spring} + \mathbf{F}_{beam} + \mathbf{F}_{target}\\
\end{equation}

Once the total force from Eq.(\ref{deform_force}) has been calculated, it needs to be spread from the Lagrangian frame to the Eulerian grid. This is achieved through an integral transform with a delta function kernel,  
\begin{equation}
\label{body_force} {\bf{f}}({\bf{x}},t)= \int \bf{F}(s,t) \delta({\bf{x}}-{\bf{X}}(s,t)) ds. \\ 
\end{equation}
Similarly, to interpolate the local fluid velocity onto the Lagrangian mesh, the same delta function transform is used, 
\begin{equation}
\label{struc_velocity} {\bf{U}}(s,t)  = \frac{\partial {\bf{X}} }{\partial t}(s,t) =  \int {\bf{u}}( {\bf{x}}, t) \delta ({\bf{x}}-{\bf{X}}(s,t)) d{\bf{x}}. \\
\end{equation}
Eqs.(\ref{body_force}) and (\ref{struc_velocity}) describe the coupling between the immersed boundary and the fluid, e.g., the communication between the Lagrangian framework and Eulerian framework. The delta functions in these equations make up the heart of the IBM, as they are used to spread and interpolate dynamic quantities between the fluid grid and elastic structure, e.g., forces and velocity. The quantity ${\bf{X}}(s,t)$ gives the position in Cartesian coordinates of the elastic structure at local material point, $s$, and time $t$. In approximating these integral transforms, a discretized and regularized delta function, $\delta_h(\mathbf{x})$ \cite{Peskin:2002}, is used, 
\begin{equation}
\label{delta_h} \delta_h(\mathbf{x}) = \frac{1}{h^2} \phi\left(\frac{x}{h}\right) \phi\left(\frac{y}{h}\right),
\end{equation}
where $\phi(r)$ is defined as

\begin{equation}
\label{delta_phi} \phi(r) = \left\{ \begin{array}{c} \frac{1}{4}\left[1 + \cos\left(\frac{\pi r}{2}\right)\right] \ \ \ \ \ |r|\leq 2 \\
$ $\\
\ \ \ \ \ \ \ \ \ \ 0 \ \ \ \ \ \  \ \ \ \ \ \ \ \  \mbox{otherwise}. \end{array} \right.
\end{equation}


%
\subsection{Numerical Algorithm}
As stated above, we impose periodic boundary conditions on the rectangular domain. To solve Eqs. (\ref{momentum}), (\ref{incompressible}),(\ref{body_force}) and (\ref{struc_velocity}) we need to update the velocity, pressure, position of the boundary, and force acting on the boundary at time $n+1$ using data from time $n$. IBM does this in the following steps \cite{Peskin:2002}:
\begin{itemize}
\item[\emph{Step 1:}] Find the force density, ${\bf{F}}^{n}$ on the immersed boundary, from the current boundary configuration, ${\bf{X}}^{n}$.
\item[\emph{Step 2:}] Use Eq.(\ref{body_force}) to spread this boundary force from the curvilinear mesh to nearby fluid lattice points.
\item[\emph{Step 3:}] Solve the Navier-Stokes equations, Eqs.(\ref{momentum}) and (\ref{incompressible}), on the Eulerian domain. In doing so, we are updating ${\bf{u}}^{n+1}$ and $p^{n+1}$ from ${\bf{u}}^{n}$ and ${\bf{f}}^{n}$. Note: because of the periodic boundary conditions on our computational domain, we can easily use the Fast Fourier Transform (FFT) \cite{Cooley:1965,Press:1992}, to solve for these updates at an accelerated rate.
\item[\emph{Step 4:}] Update the material positions, ${\bf{X}}^{n+1}$,  using the local fluid velocities, ${\bf{U}}^{n+1}$, using ${\bf{u}}^{n+1}$ and Eq.(\ref{struc_velocity}).
\end{itemize}

The above steps outline the process used by the IBM to update the positions and velocities of both the fluid and elastic structure. We note that since we are using IBAMR that additional steps are used for adaptive mesh refinement. A more detailed discussion of IBM and IBAMR are found in \cite{Peskin:2002} and  \cite{BGriffithIBAMR}, respectively.

%
%

\subsection{Model Geometry}

\begin{figure}
\begin{subfigure}{\linewidth}
\centering
\includegraphics[scale=0.78]{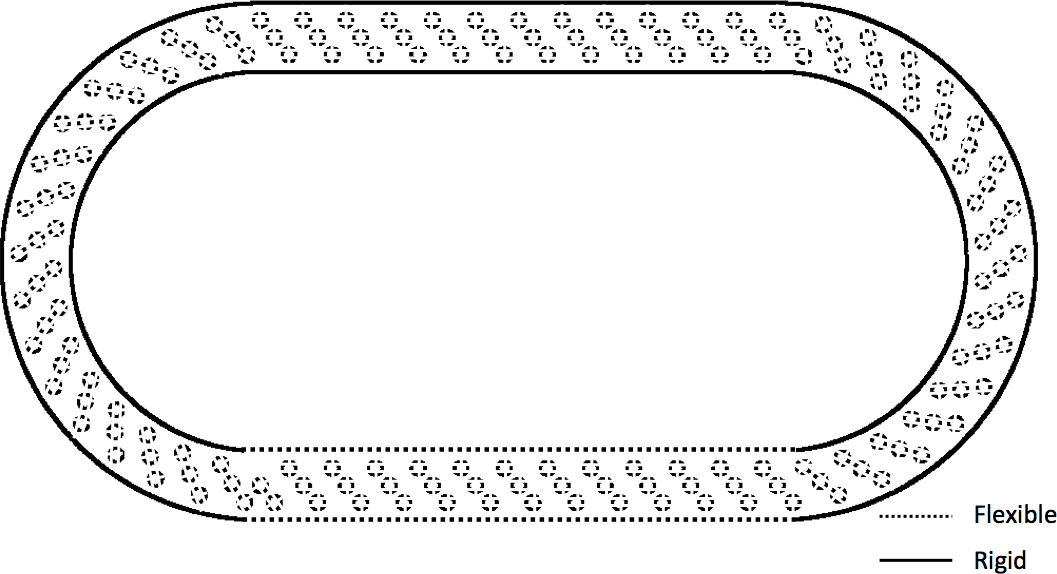}
\caption{}
\label{geometry}
\end{subfigure}\\[1ex]
\begin{subfigure}{\linewidth}
\centering
\includegraphics[scale=0.8]{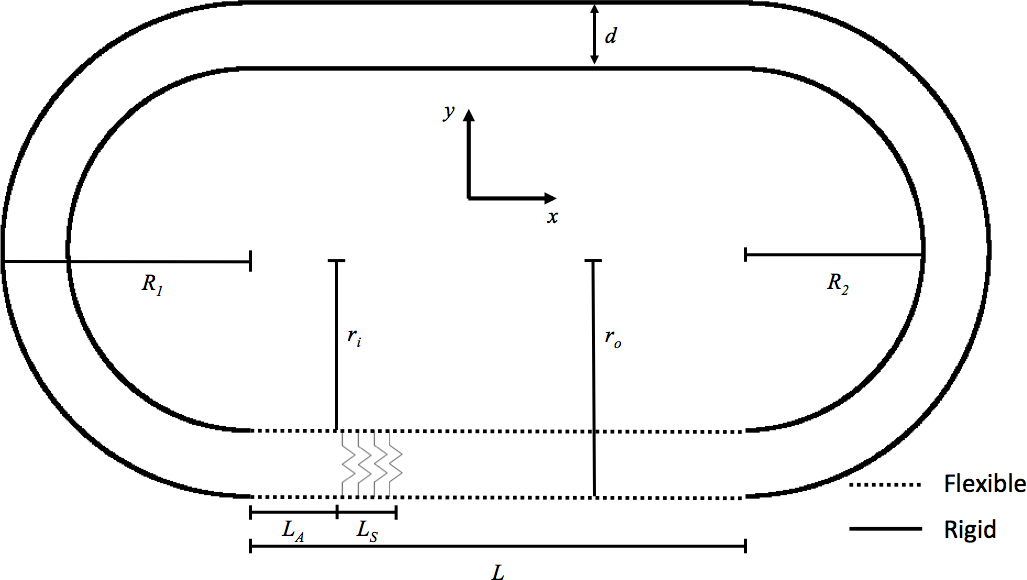}
\caption{}
\label{geometry_labeled}
\end{subfigure}
\caption{(a) illustrates the racetrack geometry, which is held rigid except for the bottom of the tube which is flexible. It also includes flexible blood cells, here illustrating the initial position for a volume fraction of $15\%$. (b) depicts the geometrical features of the racetrack.}
\label{Model_Geometry}
\end{figure}

We numerically model a $2D$ closed racetrack where the walls of the tube are modeled as $1D$ fibers. The closed tube is composed of two straight portions, of equal length, connected by two half circles, of equal inner and equal outer radii. The tube, or racetrack, has uniform diameter throughout. The geometry of the racetrack is given in Figure \ref{Model_Geometry}. 


This study goes beyond previous work \cite{Baird:2014,BairdThesis:2014,Jung:2001} through the addition of deformable blood cells, composed of springs connecting adjacent and opposite side Lagrangian nodes. The blood cells are modeled circular, in agreement with \emph{in vivo} imaging illustrating their spherical geometry in embryonic blood \cite{Maes:2011}, rather than biconcave \cite{Crowl:2009}. 

All of the mock blood cells in our simulations have the same radii. The diameter of the blood cells was set to $d/5$ \cite{Maes:2008}. The flexible cells were modeled via attaching springs between adjacent Lagrangian points for each cell, i.e. beams and target points are not used. The geometry of the heart tube with mock blood cells is illustrated in Figure \ref{geometry} with all parameter values listed in Table\ref{GeometryParams}. It is important to note that everywhere within our rectangular domain, the fluid has \emph{constant} density $\rho$ and viscosity $\mu$, even within our elastic structures.

\begin{table}[]
\centering
\begin{tabular}{|c|c|}
\hline
{\bf{Parameter}}       & {\bf{Value}} \\
\hline
$d$ & 1 \\
$R_1=r_o$ & 3.75 \\
$R_2=r_i$ & 2.75 \\
$L$ & 7.5 \\
$L_A$ & 0.9375 \\
$L_S$ & 0.75 \\
$r_C$ & 0.1 \\
$\frac{h_T}{b_V}$ & \{0, 0.1, 0.2, \ldots, 0.8\}\\
\hline
\end{tabular}
\caption{Geometric parameters used in the numerical experiments.}
\label{GeometryParams}
\end{table}


%
%
%
%

\subsubsection{DSP Model}
\label{DSP_Model_Section}

In the DSP model, the straight portion on the bottom of the racetrack geometry is flexible, e.g., is composed of beams and springs and is not tethered to target points. All other sides of the tube are held nearly rigid in a fixed position using target points.There are also springs attached over a finite actuation region from the inner to outer boundary in the bottom elastic section of the tube. These springs are used to actuate the tube, modeling DSP.  We model the action of ``muscles" with linear springs, whose resting lengths change in time. These springs are attached between the inner and outer Lagrangian boundaries of the heart-tube. 

Rather than attaching these muscles between all points within this region, we choose a region that is $10\%$ of the length of the flat portion, $L_S=L/10$, which is also translated a distance of $L_A=L/8$ along the tube from the beginning of the flat portion from the left. This model was selected since traditional DSP only assumes an off-center region of active contraction. The resting lengths of these springs were changed according to 
\begin{equation}
\label{restinglength} R_L(s,t)  = d\Big(1 - \frac{8.5}{10} \Big|  \sin( 2.3\pi t) \Big| \Big) \\
\end{equation}


%
%
%
%

\subsubsection{Peristalsis Model}

A prescribed motion of the actuation region along the bottom straight portion of the tube is used to drive peristalsis. To permit volume conservation in the closed racetrack, the top straight section of the racetrack is modeled using springs and beams and is allowed to expand. The rest of the racetrack geometry composed of target points is held nearly rigid, similarly to Section \ref{DSP_Model_Section}.  There are also springs connecting the outer and inner layer of the top of the tube for additional support. The peristaltic wave of contraction is prescribed by interpolating between multiple positions as described below. 

\begin{figure}
\centering
    \includegraphics[scale=0.375]{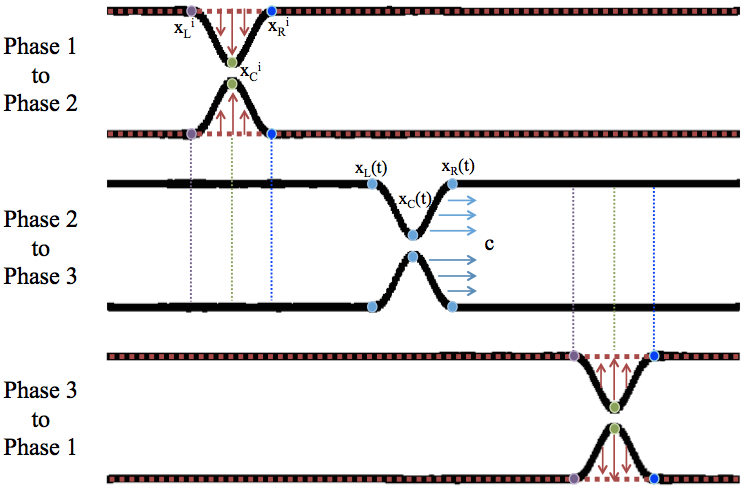}
    \caption{Interpolation phases for the traveling contraction wave along the bottom portion of the racetrack geometry. From Phase 1 (straight red tube) to Phase 2, the tube gets pinched on the left side. From Phase 2 to Phase 3, the occlusive pinch travels down the tube at speed $c$. From Phase 3, the pinch is released and goes back to the straight tube (in red).}
    \label{Phases}
\end{figure} 

Phase 1 is defined by the position of the relaxed, straight tube. Phase 2 is defined as a fully pinched tube at the initial position of contraction. Phase 3 is defined as a full pinched tube at the end of the peristaltic wave. The initial contraction (pinching) of the tube was prescribed by interpolating between Phase 1 and Phase 2. Similarly, the contractile release was performed by interpolating back between Phase 3 and Phase 1. This is illustrated in Figure \ref{Phases}. The traveling wave of contraction was performed by translating the pinch along the length of the contractile section of the tube.  

\begin{table}[]
\centering
\begin{tabular}{|c|c|}
\hline
\hline
Parameter &  Time    \\
\hline
$T$ & $0.435$ \\
$T_1$ & $0.025\times T$   \\ 
$T_2$ & $0.95\times T$   \\
$T_3$ & $0.025\times T$   \\
\hline 
\hline
\end{tabular}
\caption{Table of temporal parameters for the prescribed peristaltic wave.}
\label{InterpTime}
\end{table}

The motion motion of the actual immersed boundary is driven by changing the position of the target points, which are tethered to each immersed boundary point along the racetrack. The times of each phase (contraction, translation, and relaxation) are seen in Table \ref{InterpTime}, where $T$ is the period of one complete peristaltic wave. The following function was used to induce the traveling peristaltic wave between Phases $2$ and $3$,
\begin{equation}
\label{translation_wave_eqn} X_{target}  = \left\{ \begin{array}{l}
 \pm \tilde{A} (x-x_{L}(t))^2(x_{R}(t)-x)^2 \exp{- \frac{(x-x_{C}(t))^2}{(0.5w)^2} } - R_{o/i},\ \ \ \ x\in[x_{L}(t),x_{R}(t)]  \\
0 \ \ \ \ \ \ \ \ \ \ \ \ \ \ \ \ \ \ \ \ \ \ \ \ \ \ \ \ \ \ \ \ \ \ \ \ \ \ \ \ \ \ \ \ \ \ \ \ \ \ \ \ \ \ \ \ \ \ \ \ \ \ \ \ \ \ \ \ \  \mbox{elsewhere} \end{array} \right.,
\end{equation}
and  
\begin{align*}
    x_L(t) &= x_{L}^{i} + c(t-T_1),\\
    x_C(t) &= x_{C}^{i} + c(t-T_1),\\
    x_R(t) &= x_{R}^{i} + c(t-T_1),\\
    c&= -\frac{2x_{C}^i}{T_2},\\
    \tilde{A}&=850.0,\\
\end{align*}
where $x_{L}^{i}, x_{C}^{i},$ and $x_{R}^{i}$ are the left-most, center, and right-most points associated with the first pinch. These points are illustrated in Figure \ref{Phases}. The parameters $c$ and $\tilde{A}$ are the wave speed and amplitude, respectively. $\pm \tilde{A}$ and $R_{o/i}$ correspond to the bottom and the top wall of the tube, respectively.

%
%
%
%

\subsubsection{Determining Biologically Relevant Parameter Values}

To determine the lower range of $Wo$ within the heart tube, we take characteristic values for zebrafish embryonic hearts between $26$ and $30$ hpf and match our non-dimensionless model parameters accordingly. The characteristic frequency, $f_{zf}$ was measured \textit{in vivo}, and the characteristic length, $L_{zf}$, was taken as the diameter of the heart tube. The $Wo$ was then calculated as 
\begin{equation}
Wo = L_{zf}\ \sqrt{ \frac{2\pi\cdot f_{zf}\cdot \rho_{zf} }{ \mu_{zf} }  }  = 0.15,
\end{equation}
\noindent where $f_{zf} = 2.2\ s^{-1}$ \cite{Malone:2007}, $\rho_{zf} = 1025$ $kg/m^3$ \cite{Santhanakrishnan:2011}, $\mu_{zf} = 0.0015$ $kg / (m\cdot s)$ \cite{Mohammed:2011}, and $L_{zf} = 0.05\ mm$ \cite{Baird:2014}. The occlusion ratio is assumed to be $occ = 0.85$ \cite{Maes:2008}. We take the characteristic velocity to be $V_{pump} = f_{zf}\cdot occ \cdot\frac{L_{zf}}{2} = 0.047\ mm/s$. The dimensionless frequency may then be calculated as
\begin{equation}
\tilde{f}=\frac{L_{zf}}{V_{pump}}\cdot f_{zf} = 2.3.
\end{equation}

For the mathematical model, the parameters values were chosen to keep the dimensionless frequency fixed at $2.3$. The $Wo$ was varied by changing the dynamic viscosity, $\mu$. For the simulations, the $Wo_{sim}$ is calculated using a characteristic length of $d$, the width of the tube. The simulations were performed for $Wo_{sim} =  \{0.2,0.4,0.6,0.8,0.9,1.0,1.5,2,\ldots,9,10,15,20,30\}$. Note that the higher end of these values describe a fully inertial regime which may be outside of what is found in nature. The stiffness of the target tethering points were chosen to minimize the deformations of the boundary, i.e, to keep it rigid, and were directly correlated to $Wo$. The motivation for the wide range of $Wo$ considered is that we want to compare parameter values relevant to other types of tubular hearts, such as salps, tunicates, and insects. We also want to compare our results to the $Wo$ range considered in most previous DSP studies, (typically $Wo>1$). The other mechanical parameters were chosen to allow deformation and reexpansion of the heart tube on relevant timescales. Our parameter choices are given in Table \ref{MechanicalParams}, where they have been non-dimensionalized, using the following relations for springs (and target points) and beams, respectively,
\begin{eqnarray}
\label{nondom:spring} \tilde{k}_{spring/target} &=& \frac{ k_{spring/target} }{\mu V_{pump}} \\
\label{nondom:beam} \tilde{k}_{beam} &=& \frac{ k_{beam} }{\mu_{zf} V_{pump} L_{zf}^2}.
\end{eqnarray}
%

\begin{table}[]
\centering
\begin{tabular}{ | c | c | c | c |}
\hline
{\bf{Mechanical Parameters}}     & {\bf{Symbol}}       & {\bf{DSP Value}} & {\bf{Peristalsis Value}}  \\   \hline
Stretching Stiffness of the Tube                      & $\tilde{k}_{s_{tube}}$      & $1.0e8$          & $2.5e7$\\  \hline
Stretching Stiffness of Springs Across Tube& $\tilde{k}_{s_{btwn}}$     & $1.75e3$        & $5.0e3$\\ \hline
Stretching Stiffness of Target Points             & $\tilde{k}_{target}$           & $1.0e4$          & $1.0e6$ \\ \hline
Bending Coefficient of the Tube                    & $\tilde{k}_{beam}$          & $1.0e2$           & $1.0e10$ \\  \hline
Stretching Stiffness of Blood Cells                &   $\tilde{k}_{s_{cell}}$      & $2.15e6$         & $2.15e6$\\ \hline
\end{tabular}
\caption{Table of mechanical parameters used in the computational model. Note that $\tilde{k}_{s_{btwn}}$ gives the stiffness coefficient of the actuating springs in the DSP model, while it describes the stiffness coefficients of springs connecting the outer and inner layer of the top of the tube in the peristalsis model.}
\label{MechanicalParams}
\end{table}

\section{Results}

\begin{figure}
\centering
\includegraphics[scale=1.2]{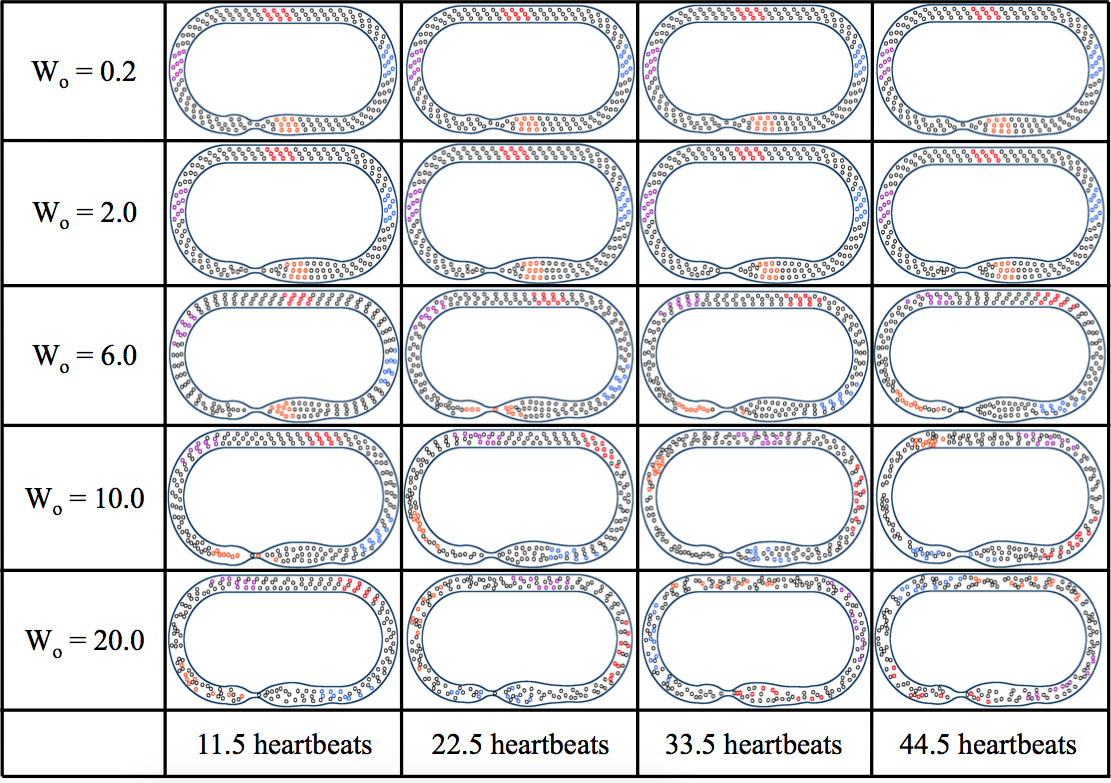}
\caption{A comparison of simulations with different Womersley Number, $Wo=\{0.2,2.0,6.0,10.0,20.0\}$, but same amount of blood cells, $VF=15\%$. The images were taken after at $11.5, 22.5, 33.5,$ and $44.5$ heartbeats during the simulations. In the case of $Wo=0.2$ and $Wo=2.0$, there is no visual transport for the mock blood cells; however, there is clear transport when $Wo\geq 6.0$. Moreover, in the cases when $Wo\geq 6.0$, the hematrocrit begins to clump together, rather than move uniformly throughout the tube.}
\label{TimeSlices}
\end{figure}

\begin{figure}
    \centering
    \includegraphics[scale=0.4]{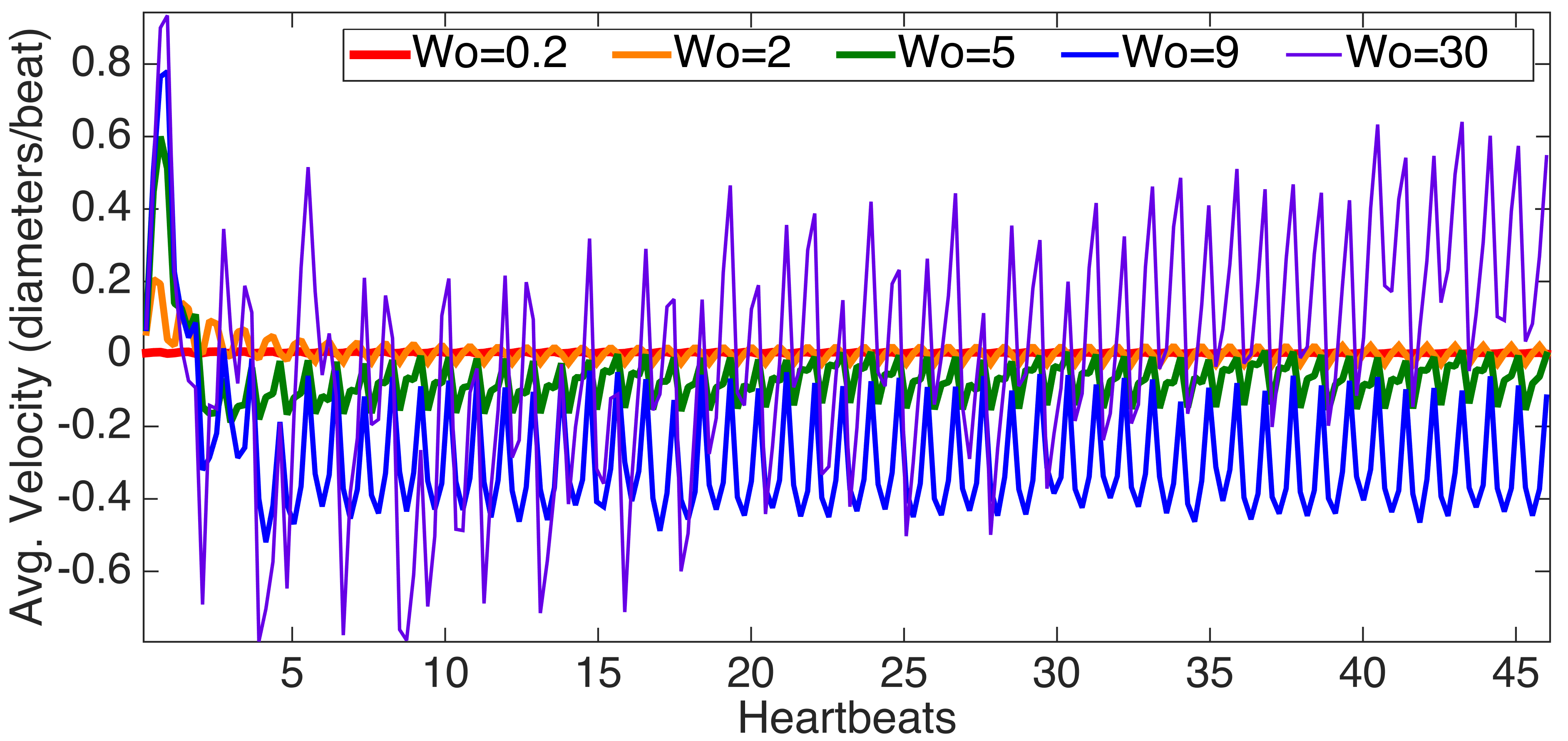}
    \caption{A comparison of the spatially averaged velocity vs. time over the course of the simulation, for five cases with uniform hematocrit ($VF=15\%$), but varying Womersley Number, $Wo=\{0.2,2,5,9,30\}$. The average velocity was spatially computed across a cross-section in the center of the top of the tube. As $Wo$ increases the amplitude of oscillations in average velocity also increases. In the biologically relevant case, $Wo=0.2$, there are slight oscillations; however, bulk net flow is insignificant.}
\label{FixVF15_VaryWo}
\end{figure}


\begin{figure}
    \begin{subfigure}[b]{0.5\textwidth}
        \centering
        \includegraphics[width=\textwidth]{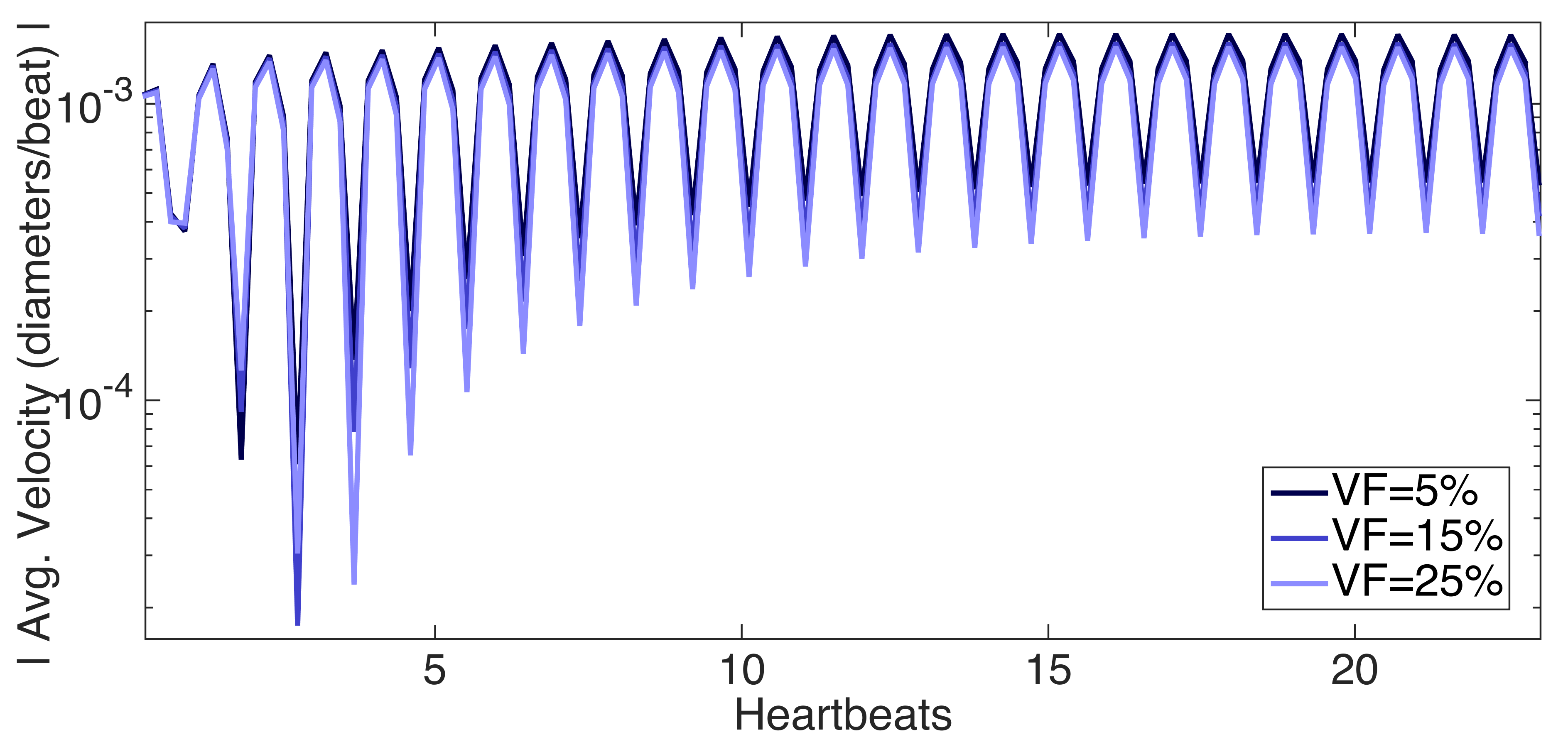} \hfill
        \caption{}
        \label{FixWo_VaryVF_Wo0pt2}
    \end{subfigure} 
    \begin{subfigure}[b]{0.5\textwidth}
        \centering
        \includegraphics[width=\textwidth]{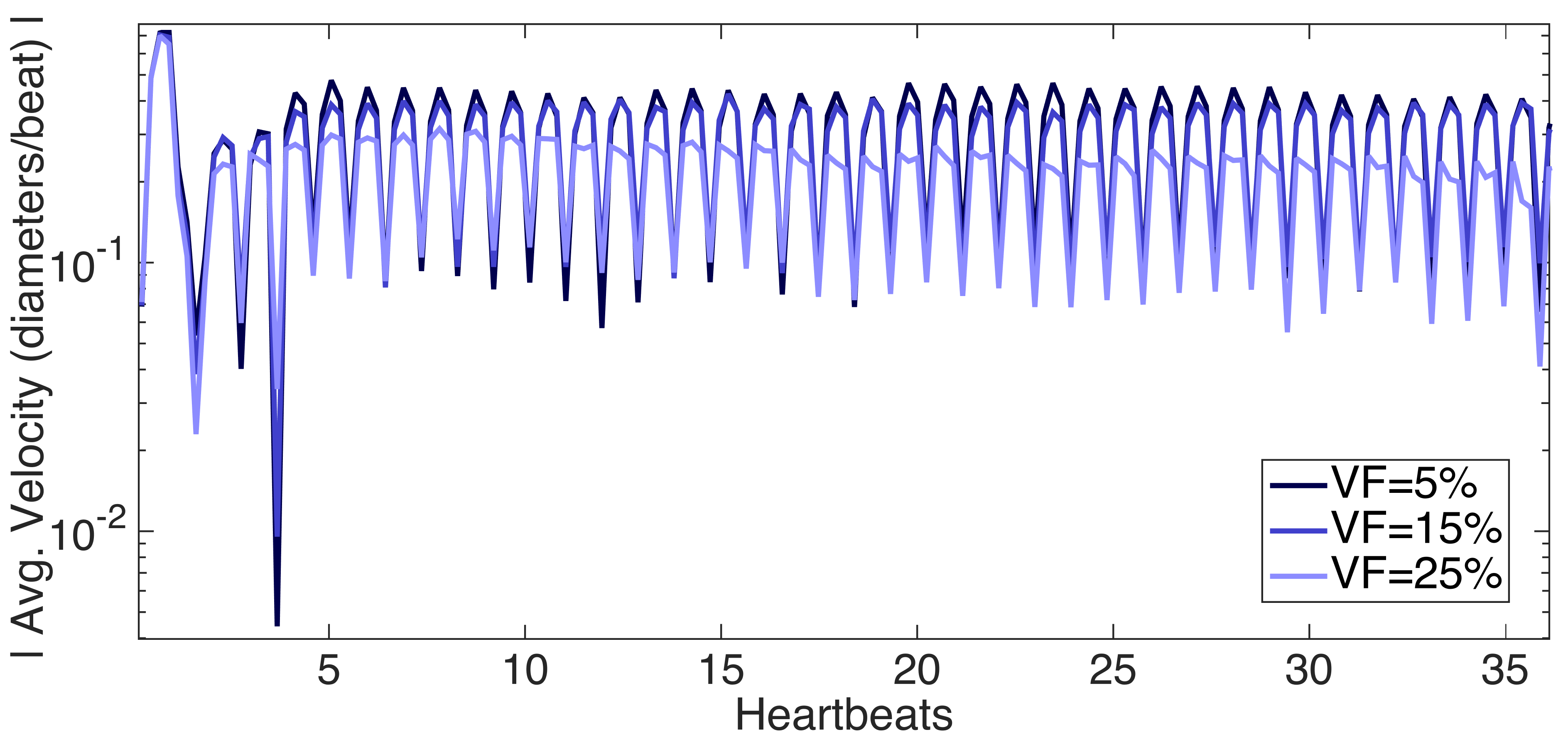} \hfill
        \caption{}
        \label{FixWo_VaryVF_Wo7}
    \end{subfigure}\\
    \begin{subfigure}[b]{0.32\textwidth}
        \includegraphics[width=\textwidth]{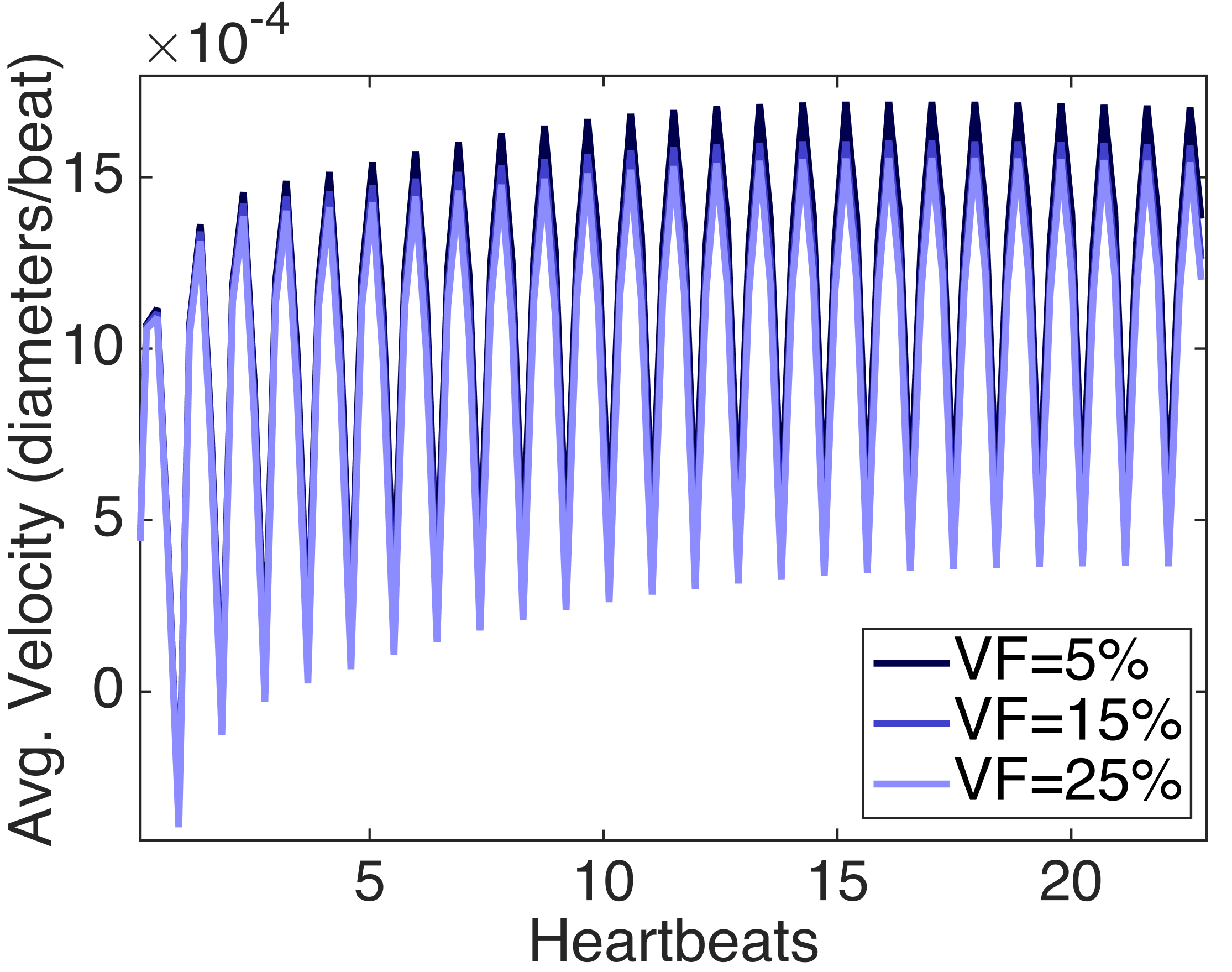}\hfill
        \caption{}
        \label{FixWo0pt2_noLog}
        \end{subfigure}
    \begin{subfigure}[b]{0.32\textwidth}
        \includegraphics[width=\textwidth]{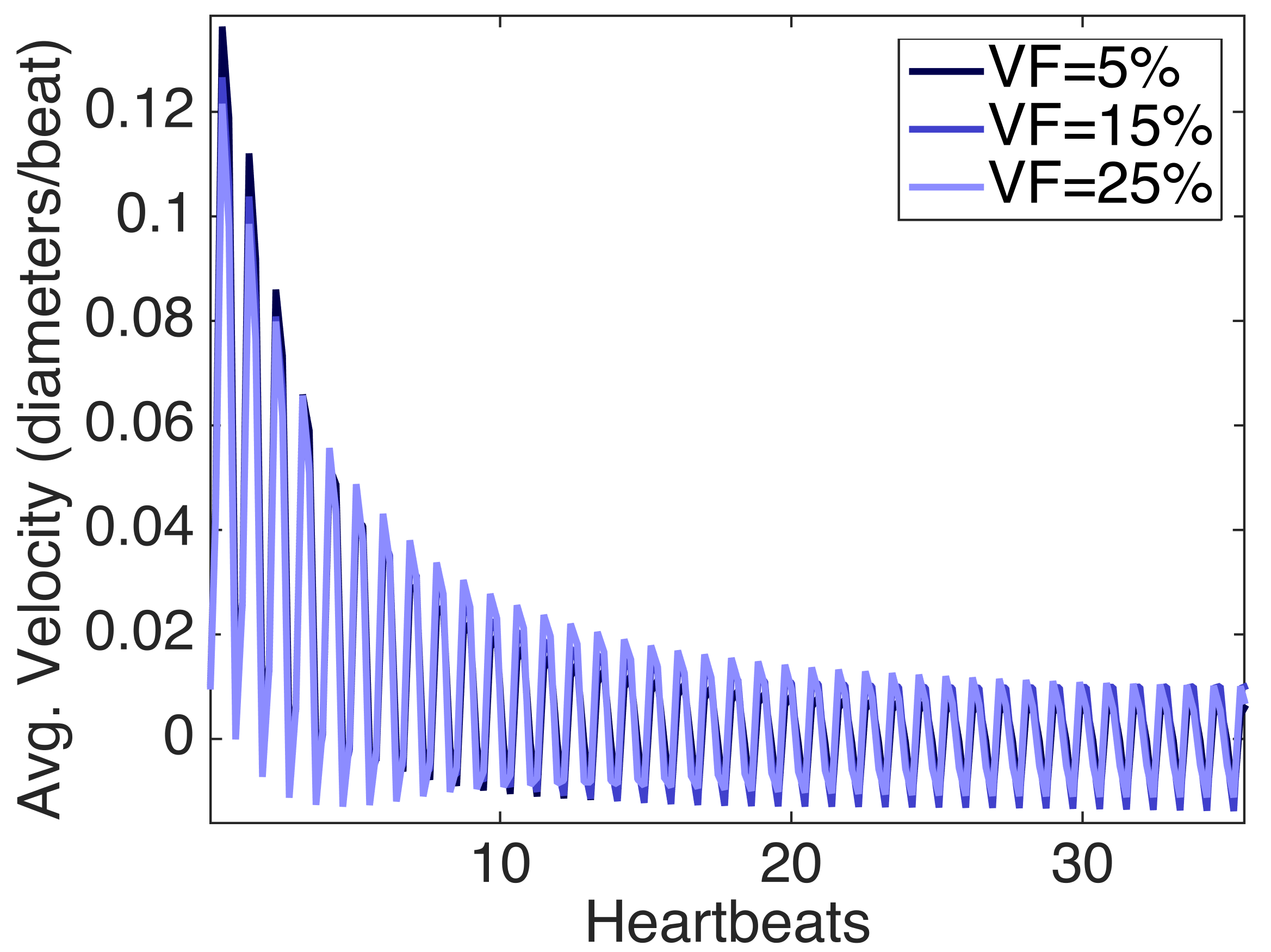}\hfill
        \caption{}
        \label{FixWo1pt5_noLog}
        \end{subfigure}
    \begin{subfigure}[b]{0.32\textwidth}
        \includegraphics[width=\textwidth]{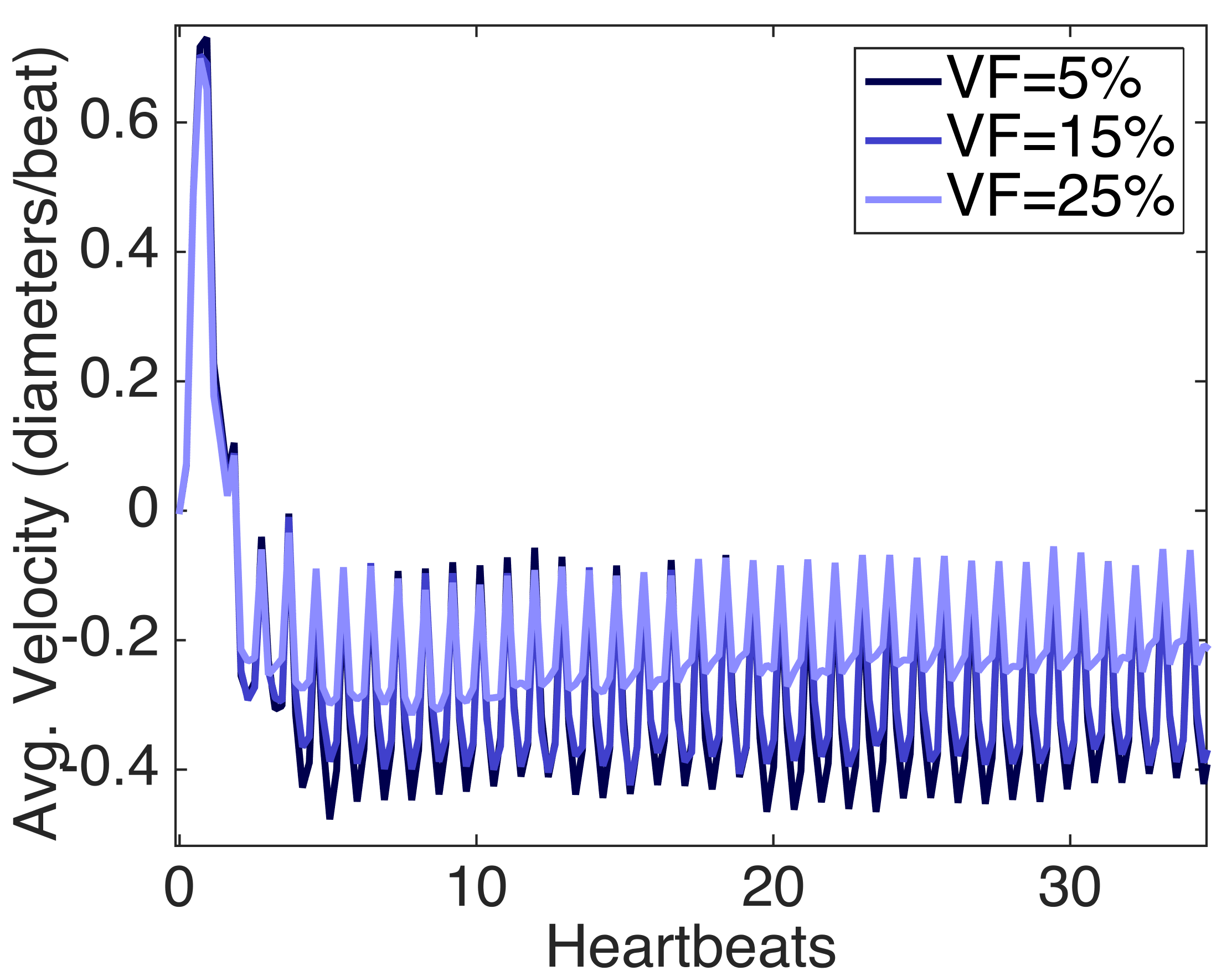}
        \caption{}
        \label{FixWo7_noLog}
    \end{subfigure}
\caption{A comparison of the spatially averaged velocity vs. time over the course of the simulation, for varying hematocrit,$VF=\{5\%,15\%,25\%\}$, for three different Womsersley Numbers, $Wo=0.2$ [(\ref{FixWo_VaryVF_Wo0pt2}), (\ref{FixWo0pt2_noLog})], $Wo=1.5$ (\ref{FixWo1pt5_noLog}), and $Wo=7.0$ [(\ref{FixWo_VaryVF_Wo7}),(\ref{FixWo7_noLog})]. The average velocity was spatially computed across a cross-section in the center of the top of the tube. [(\ref{FixWo_VaryVF_Wo0pt2}),(\ref{FixWo_VaryVF_Wo7})] illustrate how similar the waveforms are for varying volume fractions for $Wo$, $0.2$ and $7$, respectively. [(\ref{FixWo0pt2_noLog}),(\ref{FixWo1pt5_noLog}),(\ref{FixWo7_noLog})] give the average velocities, in diameters/heartbeat, over the course of the simulation, in heartbeats.}
\label{FixWo_VaryVF}
\end{figure}


\begin{figure}
\begin{subfigure}[b]{0.5\textwidth}
\centering
\includegraphics[width=\textwidth]{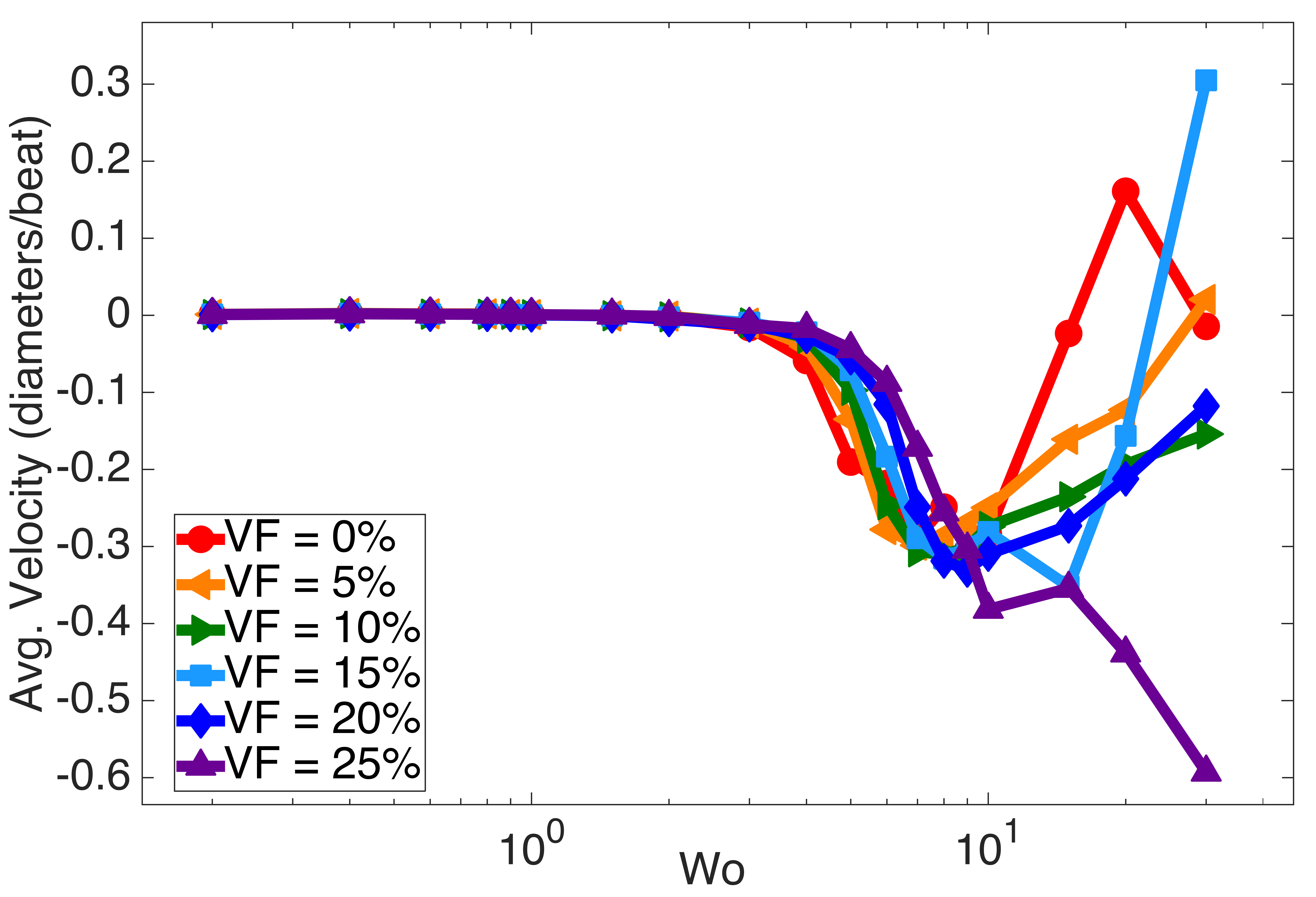}
\caption{}
\label{AvgV}
\end{subfigure} 
\begin{subfigure}[b]{0.5\textwidth}
\centering
\includegraphics[width=\textwidth]{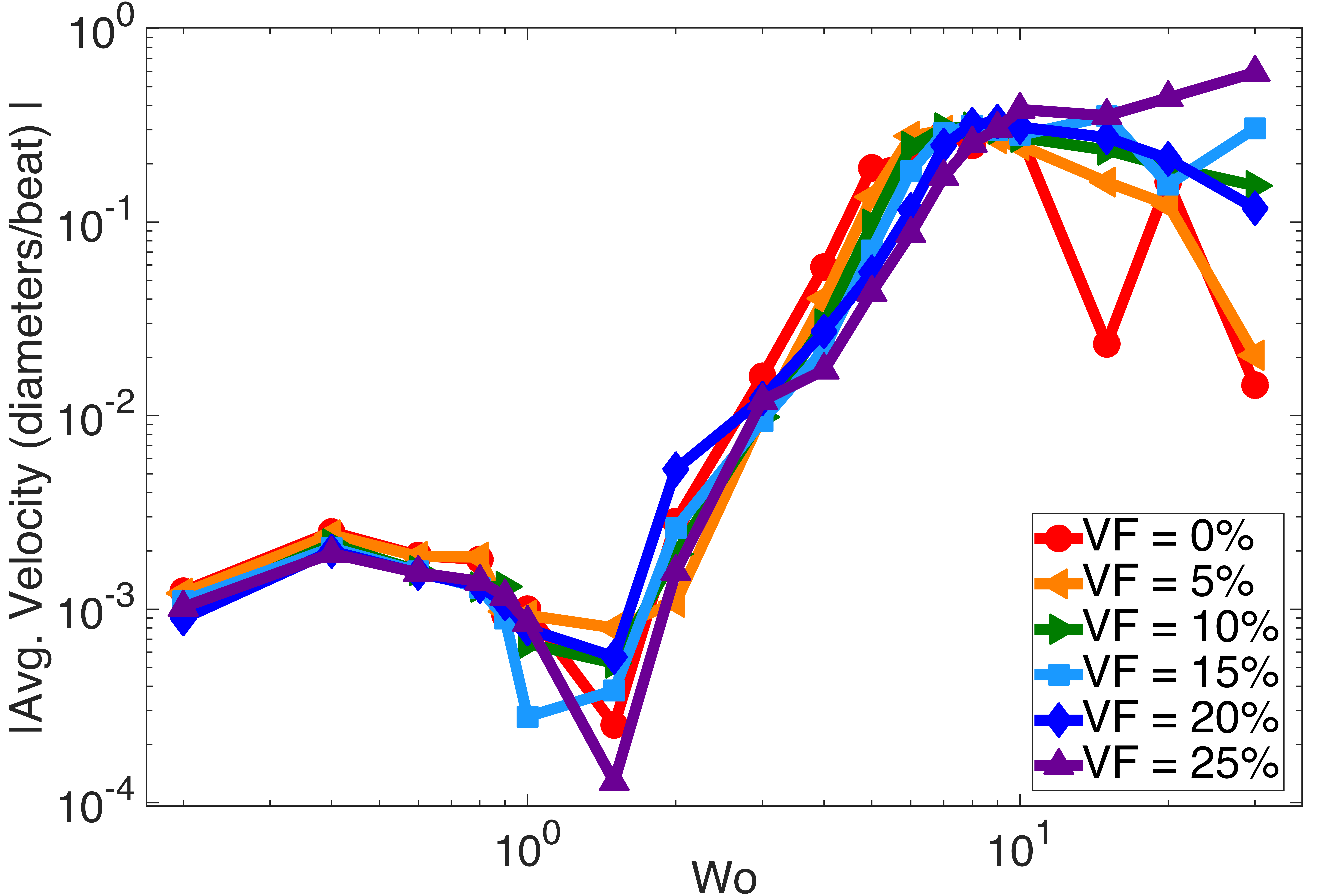}
\caption{}
\label{AbsAvgV}
\end{subfigure}
\caption{\ref{AvgV} shows the spatially- and temporally-averaged velocity for each simulation vs. Womerseley Number for a hematocrit range of $[0\%,25\%]$. \ref{AbsAvgV} shows the spatially- and temporally-averaged magnitude of velocity vs. $Wo$ for a hematocrit range of $[0\%,25\%]$.}
\label{AvgV_Plots}
\end{figure}

In this paper, we present simulations of dynamic suction pumping and peristalsis within a closed racetrack containing flexible blood cells of varying volume fractions but uniform geometry. The simulations were run for a range of Womersley Number, $Wo\in[0.1,30]$, and hematocrit, $VF=\{0\%,5\%,10\%,15\%,20\%,25\%\}$. Examples of the locations of the blood cells and boundaries at different points in time are seen in Figure \ref{TimeSlices}.

%
%
%
%

\subsection{DSP Results}

Figure \ref{TimeSlices} shows snapshots from simulations of DSP at five different Womersley numbers, $Wo=\{0.2,2.0,6.0,10.0,20.0\}$, where volume fraction is held constant at $VF=15\%$. The images were taken after at $11.5, 22.5, 33.5,$ and $44.5$ heartbeats. In the cases of $Wo=0.2$ thru $Wo=2.0$, there is no significant net transport for the mock blood cells as evidenced by the negligible movement of the blood cells (note the color coding of blood cells in each quadrant). There is, however, clear transport when $Wo\geq 6.0$. Moreover, in the cases when $Wo\geq 6.0$, the blood cells begin to clump together, rather than move uniformly throughout the tube.

Keeping the volume fraction constant, we compared the spatially-averaged velocity across a cross-section in the center of the top of the tube, for three different Womersley Numbers, $Wo\in\{0.2,2.0,5.0,9.0,30.0\}$. Note that deformations of this section of the tube are negligble such that the average velocity is directly proportional to the volumetric flow rate. Figure \ref{FixVF15_VaryWo} illustrates this for the case of $VF=15\%$. From the figure, it is evident that the lower $Wo$ case induces less net flow than the other two higher $Wo$ cases. However, we can also deduce that the direction of flow is a non-linear function of $Wo$. Note that for $Wo=9.0$, flow is moving in the opposite direction to that of the $Wo=20.0$ case.  

Moreover, an example comparison of the spatially-averaged velocity vs. time for three different volume fractions, $VF=\{5\%,15\%,25\%\}$, for three specific Womersley numbers, $Wo=\{0.2,1.5,7.0\}$, are shown in Figure \ref{FixWo_VaryVF}. Figures[(\ref{FixWo_VaryVF_Wo0pt2}),(\ref{FixWo_VaryVF_Wo7})] show the similarity of the waveforms  illustrating little effect of blood cells on bulk flow patterns, for all three $Wo$. [(\ref{FixWo0pt2_noLog}),(\ref{FixWo1pt5_noLog}),(\ref{FixWo7_noLog})] give the spatially averaged velocities (in diameters/heartbeat) vs. time over the course of the simulation. Time is given in number of heartbeats. It is interesting to note that in both the $Wo=1.5$ and $Wo=7.0$ cases, the average velocities are greatest at the beginning of the simulation and then asymptotically decrease until they reach a periodic cycle much lower than the beginning transient velocities. In the $Wo=0.2$ case, the average velocities asymptotically increase until they reach a periodic cycle. It is clear the maximal flow rates in both the $Wo=0.2$ and $Wo=1.5$ cases are multiple orders of magnitude below one diameter/heartbeat. These are well below the experimentally observed velocity of $\sim0.9$ diameters/heartbeat recorded in zebrafish \cite{Forouhar:2006}.

To quantify the effect of blood cells further, spatially- and temporally-averaged velocities for various $Wo$ and hematocrits were compared. This is illustrated in Figure \ref{AvgV_Plots}. From Figure \ref{AvgV}, it is clear that flow rates are a non-linear function of $Wo$. The case with zero hematocrit is in agreement with previous results reported in \cite{Baird:2014}. Moreover, the addition of hematocrit does not significantly perturb flow rates for $Wo\lesssim10$, as seen in Figure \ref{AbsAvgV}. However, for $Wo\gtrsim10$, the addition of hematocrit affects flow rates in a non-linear fashion.

\subsection{Peristalsis Results}

\begin{figure}
\centering
\includegraphics[scale=1.1]{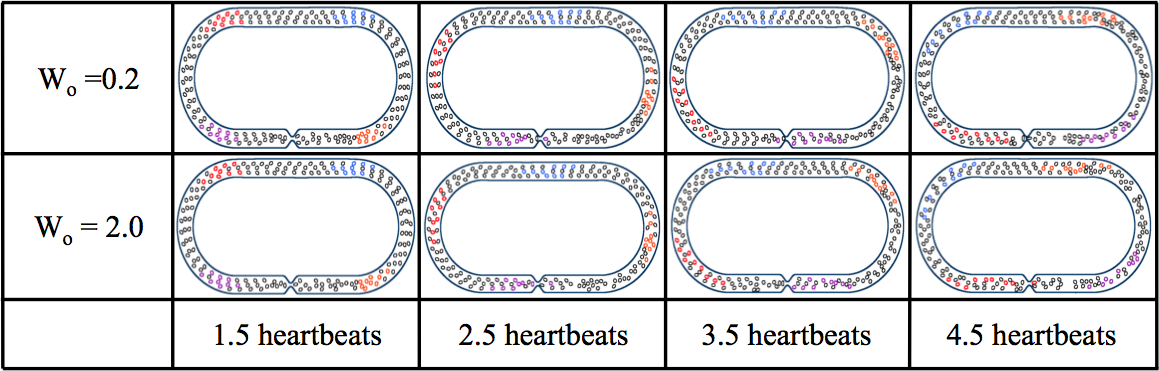}
\caption{A comparison of simulations for two different Womersley Numbers, $Wo=\{0.2,2.0\}$, but same hematocrit, $VF=15\%$. The images were taken after at $1.5, 2.5, 3.5$, and $4.5$ heartbeats during the simulations. It is clear that there is significant mixing of the blood cells with peristalsis, as the colored sections begin to mix}
\label{Peri_TimeSlices}
\end{figure}

 \begin{figure}
    \begin{subfigure}[b]{\textwidth}
        \centering
        \includegraphics[width=\textwidth]{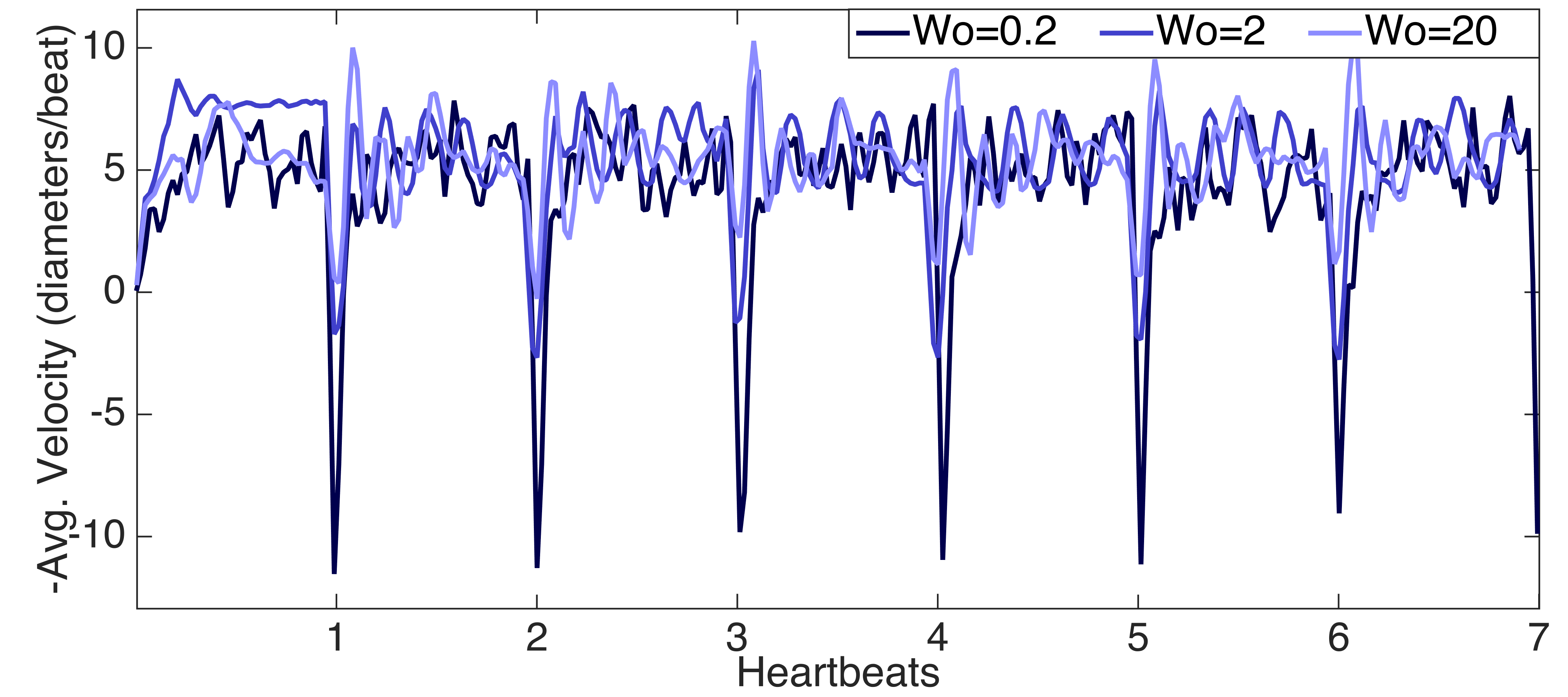} \hfill
        \caption{}
        \label{Peri_FixVF_VaryWo}
    \end{subfigure} \\
    \begin{subfigure}[b]{0.5\textwidth}
        \includegraphics[width=\textwidth]{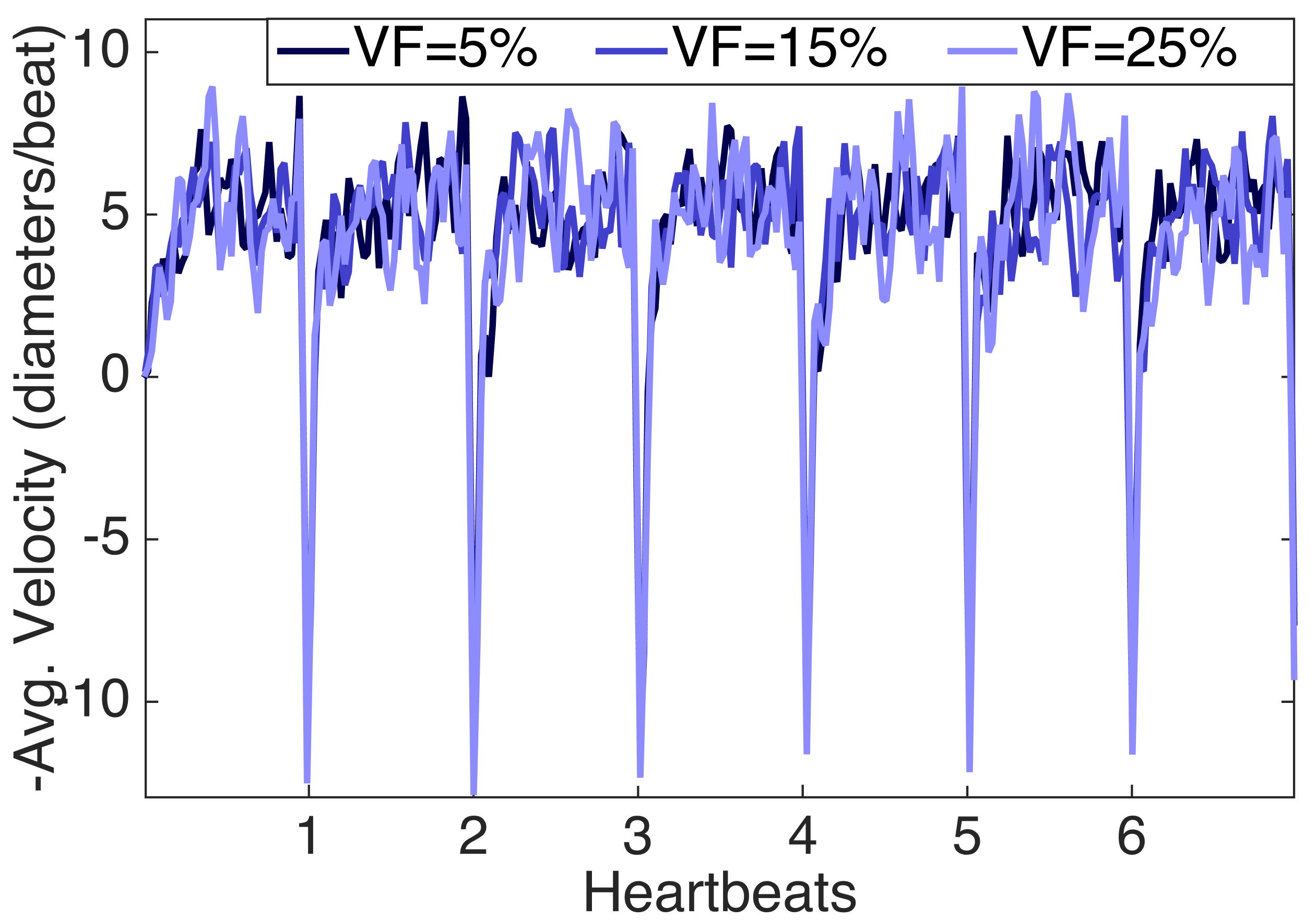}\hfill
        \caption{}
        \label{Peri_Compare_Wo0pt2_Vary_VF}
        \end{subfigure}
    \begin{subfigure}[b]{0.5\textwidth}
        \includegraphics[width=\textwidth]{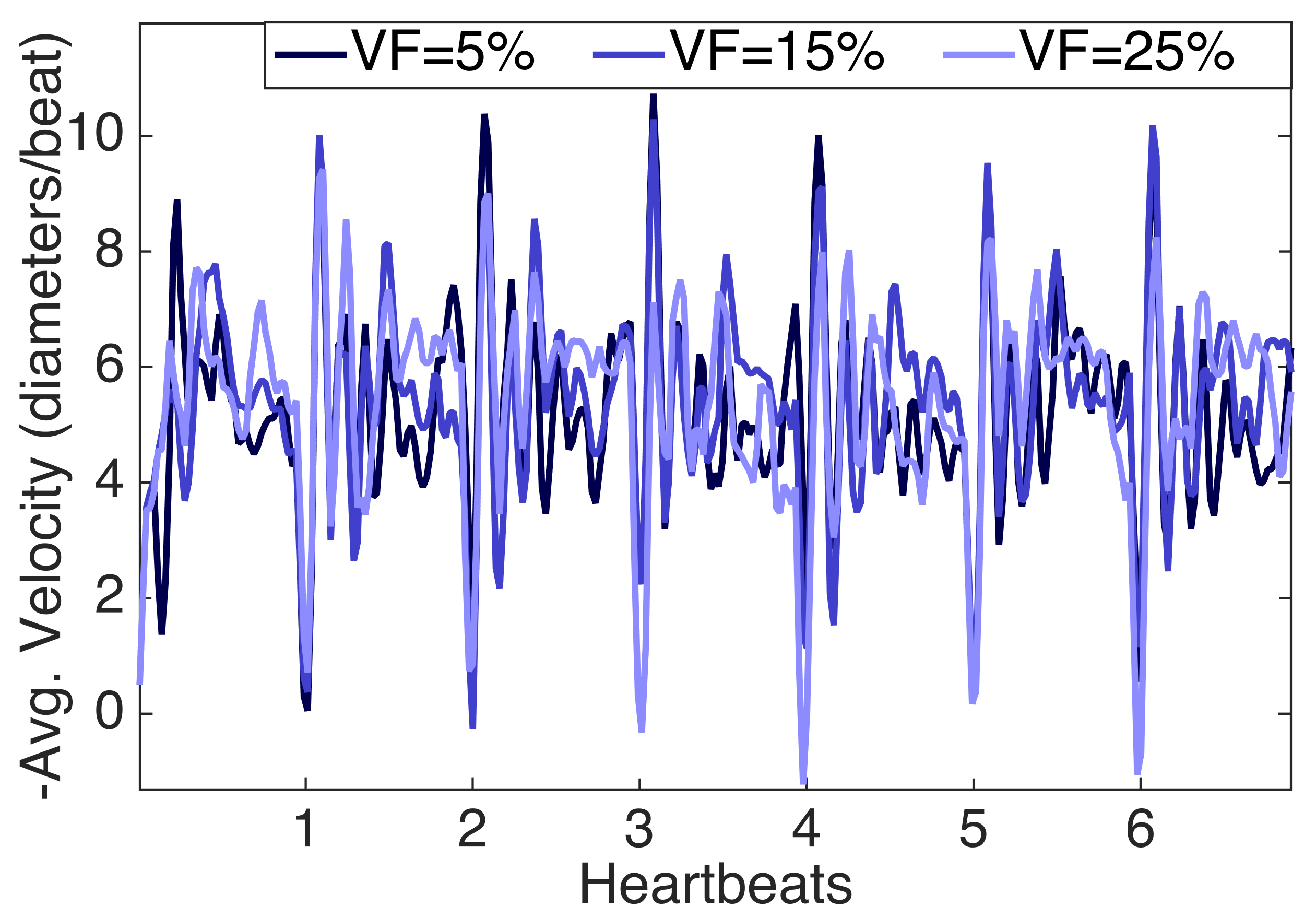}\hfill
        \caption{}
        \label{Peri_Compare_Wo20_Vary_VF}
    \end{subfigure}
\caption{A comparison of the spatially averaged velocity vs. time over the course of the simulation, for hematocrit,$VF=15\%$, for three different Womsersley Numbers, $Wo={0.2,2,20}$ is shown in (\ref{Peri_FixVF_VaryWo}). The spatially-averaged velocity was computed across a cross-section in the center of the top of the tube, given in diameters/heartbeat. (\ref{Peri_Compare_Wo0pt2_Vary_VF}) and (\ref{Peri_Compare_Wo20_Vary_VF}) give the spatially-averaged velocities for $Wo = 0.2$ and $Wo=20$, respectively, for three volume fractions, $VF={5\%,15\%,25\%}$.}
\label{Peri_Data}
\end{figure}
 
 Figure(\ref{Peri_TimeSlices}) shows snapshots from simulations for two different Womersley numbers, $Wo=\{0.2,2.0\}$, where hematocrit is held constant at $VF=15\%$. The images were taken after $1.5,2.5,3.5$, and $4.5$ heart beats during the simulations. It is clear from both simulations that there is significant bulk flow throughout the tube. Moreover, significant mixing is observed in both cases.
 
 The volume fraction was kept constant, at $VF=15\%$, in Figure(\ref{Peri_FixVF_VaryWo}) to explore the effect of scaling on bulk flow for $Wo={0.2,2.0,20.0}$. The spatially-averaged velocity across the top of the tube, given in diameters/heartbeat, is similar between all three cases of $Wo$. Furthermore the direction of flow is consistent in all cases, with bulk flow moving counterclockwise around the tube, with a sharp decrease in velocity, showing flow going in the opposite direction, at the end of each heartbeat. 
 
 Figures(\ref{Peri_Compare_Wo0pt2_Vary_VF}) and (\ref{Peri_Compare_Wo20_Vary_VF}) illustrate the effect of varying hematocrit for simulations with $Wo=0.2$ and $Wo=20$, respectively. In both cases the waveforms look similar, suggesting the addition of blood cells does not significantly affect bulk flow rates. However, we note that the sharp decrease in velocity at the end of the heartbeat is more pronounced in the $Wo=0.2$ case, than in the $Wo=20$ case.

%
%
%
%

\section{Conclusions}

In this paper, two-dimensional immersed boundary simulations were used to model dynamic suction pumping and peristalsis for a single actuation frequency over a range of Womerseley numbers and hematocrits relevant to valveless, tubular hearts. When strong net flow was generated in the tube at higher $Wo$, blood cells clumped together, and did not flow uniformly throughout the tube. The spatially- and temporally- averaged velocities across a cross-section along the top of the tube showed a non-linear relationship between net flow rates and $Wo$ for DSP. The effect of hematocrit on the net flow rate was significant for $Wo\gtrsim10$ and was nonlinear. The addition of blood cells did not enhance the weak net flows produced for $Wo<1$. These results highlight the complex dynamics governing dynamic suction pumping. 

For DSP at low $Wo$, the fluid is nearly-reversible. This reversibility may explain in part why there is little net flow in the tube for the case of DSP (a reversible motion) at $VF=0$. This result is in agreement with  \cite{Baird:2014,BairdThesis:2014}. Previous studies have shown enhanced fluid transport and animal locomotion in non-Newtonian fluids at low $Re$ and $Wo$ \cite{Lauga:2007}. Since the addition of blood cells in a Newtonian fluid makes the bulk fluid effectively non-Newtonian, it is possible that the addition of blood cells could make the flow in tubular hearts irreversible. For the parameters considered here, any such effect was negligible. 


For the case of peristalsis, flow was consistently driven around the racetrack for all $Wo$ and for all hematocrits. Similar to DSP, the addition of hematocrit did not significantly change net flow rates at low $Wo$. The addition of hematocrit also did not significantly alter the velocity waveform or the net flow at higher $Wo$. 


Although the bulk transport of fluid was not signficiantly changed, the addition of blood cells may affect the shear stresses experienced by the cardiac cells and the amount of mixing within the heart tube. The peristalsis simulations show enhanced mixing as compared to that of DSP at the same $Wo$ and $VF$. Furthermore for $Wo={0.2,2}$, peristalsis was able achieve similar levels of blood cell mixing an order of magnitude faster than the DSP simulation at $Wo=20$. These results are important when considering the role that fluid mixing and shear stress may play in cardiogenesis. 

Experimental evidence has shown blood flow, and more specifically hemodynamic forces, are essential for proper heart morphogenesis \cite{Hove:2003}. Furthermore, it is evident that there is a strongly coupled relationship between the underlying hemodynamics, cardiac electrophysiology, and activation of some genetic regulatory networks. For example, hemodynamics is thought to regulate the development of the pacemakers and the conduction of action potentials in the heart \cite{Tucker:1988, Reckova:2003}. Since there is direct feedback between the underlying electrophysiology and the flow induced by muscle contraction, changes in traveling action potentials will affect the hemodynamic forces felt at the endothelial layer, e.g., shear stress and pressure. These changes may then result in changes in gene expression via epigenetic signaling mechanisms, e.g., mechanotransduction. However, the exact pipelines that contribute to mechanotransduction are not completely understood. \cite{Weinbaum:2003}. 





\section{Acknowledgements}
The authors would like to thank Steven Vogel for conversations on scaling in various hearts.  We would also like to thank Lindsay Waldrop, Austin Baird, Jiandong Liu, Leigh Ann Samsa, and William Kier for discussions on embryonic hearts. This project was funded by NSF DMS CAREER \#1151478 awarded to L.A.M. Funding for N.A.B. was provided from an National Institutes of Health T32 grant [HL069768-14; PI, Christopher Mack].




\newpage


\bibliographystyle{spmpsci}      
\bibliography{heart}   

\begin{thebibliography}{10}
\providecommand{\url}[1]{{#1}}
\providecommand{\urlprefix}{URL }
\expandafter\ifx\csname urlstyle\endcsname\relax
  \providecommand{\doi}[1]{DOI~\discretionary{}{}{}#1}\else
  \providecommand{\doi}{DOI~\discretionary{}{}{}\begingroup
  \urlstyle{rm}\Url}\fi

\bibitem{Roubaie:2011}
Al-Roubaie, S., Jahnsen, E.D., Mohammed, M., Henderson-Toth, C., Jones, E.A.:
  Rheology of embryonic avian blood.
\newblock Am. J. Physiol. Heart Circ. Physiol. \textbf{301}(6919), 2473--2481
  (2011)

\bibitem{Auerbach:2004}
Auerbach, D., Moehring, W., Moser, M.: An analytic approach to the liebau
  problem of valveless pumping.
\newblock Cardiovascular Engineering: An International Journal \textbf{4},
  201--207 (2004)

\bibitem{Avrahami:2008}
Avrahami, I., Gharib., M.: Computational studies of resonance wave pumping in
  compliant tubes.
\newblock Journal of Fluid Mechanics \textbf{608}, 139--160 (2008)

\bibitem{Babbs:2010}
Babbs, C.: Behavior of a viscoelastic valveless pump: a simple theory with
  experimental validation.
\newblock BioMedical Engineering Online \textbf{9:42}, 19,832–19,837 (2010)

\bibitem{BairdThesis:2014}
Baird, A.J.: Modeling valveless pumping mechanisms (ph.d. thesis).
\newblock University of North Carolina at Chapel Hill \textbf{628}, 129--148
  (2014)

\bibitem{Baird:2014}
Baird, A.J., King, T., Miller, L.A.: Numerical study of scaling effects in
  peristalsis and dynamic suction pumping.
\newblock Biological Fluid Dynamics: Modeling, Computations, and Applications
  \textbf{628}, 129--148 (2014)

\bibitem{MJBerger84}
Berger, M.J., Oliger, J.: Adaptive mesh refinement for hyperbolic
  partial-differential equations.
\newblock J. Comput. Phys. \textbf{53}(3), 484--512 (1984)

\bibitem{MJBerger89}
Berger, M.J., P.Colella: Local adaptive mesh refinement for shock
  hydrodynamics.
\newblock J. Comput. Phys. \textbf{82}(1), 64--84 (1989)

\bibitem{Bringley:2008}
Bringley, T., Childress, S., Vandenberghe, N., Zhang, J.: An experimental
  investigation and a simple model of a valveless pump.
\newblock Physics of Fluids \textbf{20}, 033,602 (2008)

\bibitem{Chang:2007}
Chang, H.T., Lee, C.Y., Wen, C.Y.: Design and modeling of electromagnetic
  actuator in mems-based valveless impedance pump.
\newblock Microsystems Technologies | Micro-and Nanosystems- Information
  Storage and Processing Systems \textbf{13}, 1615--1622 (2007)

\bibitem{Cooley:1965}
Cooley, J., Tukey, J.W.: An algorithm for the machine calculation of complex
  fourier series.
\newblock Math. Comput. \textbf{19}, 297--301 (1965)

\bibitem{Crowl:2009}
Crowl, L.M., Fogelson, A.L.: Computational model of whole blood exhibiting
  lateral platelet motion induced by red blood cells.
\newblock Int. J. Numer. Meth. Biomed. Engng. \textbf{26}, 471–487 (2009)

\bibitem{Fogelson:2008}
Fogelson, A.L., Guy, R.D.: Immersed-boundary-type models of intravascular
  platelet aggregation.
\newblock Comput. Methods Appl. Mech. Engrg. \textbf{197}, 2087–2104 (2008)

\bibitem{Forouhar:2006}
Forouhar, A.S., Liebling, M., Hickerson, A., Nasiraei-Moghaddam, A., Tsai,
  H.J., Hove, J.R., Fraser, S.E., Dickinson, M.E., Gharib, M.: The embryonic
  vertebrate heart tube is a dynamic suction pump.
\newblock Science \textbf{312}(5774), 751--753 (2006)

\bibitem{BGriffithIBAMR}
Griffith, B.E.: An adaptive and distributed-memory parallel implementation of
  the immersed boundary (ib) method (2014).
\newblock \urlprefix\url{https://github.com/IBAMR/IBAMR}

\bibitem{Griffith:2007}
Griffith, B.E., Hornung, R., McQueen, D., Peskin, C.S.: An adaptive, formally
  second order accurate version of the immersed boundary method.
\newblock J. Comput. Phys. \textbf{223}, 10–49 (2007)

\bibitem{Hickerson:2005}
Hickerson, A., Rinderknecht, D., Gharib, M.: Experimental study of the behavior
  of a valveless impedance pump.
\newblock Experiments in Fluids \textbf{38}, 534--540 (2005)

\bibitem{HickersonThesis:2005}
Hickerson, A.I.: An experimental analysis of the characteristic behaviors of an
  impedance pump (ph.d. thesis).
\newblock California Institute of Technology \textbf{608}, 139--160 (2005)

\bibitem{Hieber:2008}
Hieber, S., Koumoutsakos, P.: An immersed boundary method for smoothed particle
  hydrodynamics of self-propelled swimmers.
\newblock J. Comput. Phys. \textbf{227}, 8636–8654 (2008)

\bibitem{Hove:2003}
Hove, J.R., Koster, R.W., Forouhar, A.S., Acevedo-Bolton, G., Fraser, S.E.,
  Gharib, M.: Intracardiac fluid forces are an essential epigenetic factor for
  embryonic cardiogenesis.
\newblock Nature \textbf{421}(6919), 172--177 (2003)

\bibitem{Jung:1999}
Jung, E.: Two-dimensional simulations of valveless pumping using the immersed
  boundary method (ph.d. thesis).
\newblock Courant Institute of Mathematics, New York University \textbf{608},
  139--160 (1999)

\bibitem{Jung:2001}
Jung, E., Peskin, C.: 2-d simulations of valveless pumping using immersed
  boundary methods.
\newblock SIAM Journal on Scientific Computing \textbf{23}, 19--45 (2001)

\bibitem{Kenner:2000}
Kenner, T., Moser, M., Tanev, I., Ono, K.: The liebau-effect or on the optimal
  use of energy for the circulation of blood.
\newblock Scripta Medica \textbf{73}, 9--14 (2000)

\bibitem{Kriebel:1967}
Kriebel, M.E.: Conduction velocity and intracellular action potentials of the
  tunicate heart.
\newblock The Journal of General Physiology \textbf{50}, 2097--2107 (1967)

\bibitem{Lauga:2007}
Lauga, E.: Propulsion in a viscoelastic fluid.
\newblock Phys.Fluids \textbf{19}, 083,104 (2007)

\bibitem{Lee:2008}
Lee, C.Y., Chang, H.T., Wen, C.Y.: A mems-based valveless impedance pump
  utilizing electromagnetic actuation.
\newblock Journal of Micromechanics and Microengineering \textbf{18}, 225--228
  (2008)

\bibitem{Lee:2004}
Lee, D.S., Yoon, H.C., Ko, J.S.: Fabrication and characterization of a
  bidirectional valveless peristaltic micropump and its application to a
  flow-type immunoanalysis.
\newblock Sensors and Actuators \textbf{103}, 409--415 (2004)

\bibitem{Maes:2008}
Maes, F., Chaudhry, B., Ransbeeck, P.V., Verdonck, P.: Visualization and
  modeling of flow in the embryonic heart.
\newblock IFMBE Proceedings \textbf{22}(6919), 1875--1878 (2008)

\bibitem{Maes:2011}
Maes, F., Chaudhry, B., Ransbeeck, P.V., Verdonck, P.: The pumping mechanism of
  embryonic hearts.
\newblock IFMBE Proceedings \textbf{37}, 470--473 (2011)

\bibitem{Malone:2007}
Malone, M., Sciaky, N., Stalheim, L., Klaus, H., Linney, E., Johnson, G.:
  Laser-scanning velocimetry: A confocal microscopy method for quantitative
  measurement of cardiovascular performance in zebrafish embryos and larvae.
\newblock BMC Biotechnology \textbf{7}, 40 (2007)

\bibitem{Manner:2010}
Manner, J., Wessel, A., Yelbuz, T.M.: How does the tubular embryonic heart
  work? looking for the physical mechanism generating unidirectional blood flow
  in the valveless embryonic heart tube.
\newblock Developmental Dynamics \textbf{239}, 1035--1046 (2010)

\bibitem{Manopoulos:2006}
Manopoulos, C.G., Mathioulakis, D.S., Tsangaris, S.G.: One-dimensional model of
  valveless pumping in a closed loop and a numerical solution.
\newblock Physics of Fluids \textbf{18}, 201--207 (2006)

\bibitem{Meier:2011}
Meier, J.: A novel experimental study of a valveless impedance pump for
  applications at lab-on-chip, microfluidic, and biomedical device size scales
  (ph.d. thesis).
\newblock California Institute of Technology. p. 8636–8654 (2011)

\bibitem{Miller:2004}
Miller, L.A., Peskin, C.S.: When vortices stick: an aerodynamic transition in
  tiny insect flight.
\newblock J. Exp. Biol. \textbf{207}, 3073–3088 (2004)

\bibitem{Miller:2009}
Miller, L.A., Peskin, C.S.: A computational fluid dynamics of clap and fling in
  the smallest insects.
\newblock J. Exp. Biol. \textbf{208}, 3076--3090 (2009)

\bibitem{Miller:2012}
Miller, L.A., Santhanakrishnan, A., Jones, S.K., Hamlet, C., Mertens, K., Zhu,
  L.: Reconfiguration and the reduction of vortex-induced vibrations in broad
  leaves.
\newblock J. Exp. Biol. \textbf{215}, 2716--2727 (2012)

\bibitem{Mittal:2005}
Mittal, R., Iaccarino, C.: Immersed boundary methods.
\newblock Annu. Rev. Fluid Mech. \textbf{37}, 239–261 (2005)

\bibitem{Mohammed:2011}
Mohammed, M., Roubaie, S., Jahnsen, E., Jones, E.: Drawing first blood:
  Measuring avian embryonic blood viscosity.
\newblock SURE Poster Presentation \textbf{61}, 33--45 (2011)

\bibitem{Ottsen:2003}
Ottesen, J.: Valveless pumping in a fluid-filled closed elastic tube-system:
  one-dimensional theory with experimental validation.
\newblock Journal of Mathematical Biology \textbf{46}, 309--332 (2003)

\bibitem{Peskin:1977}
Peskin, C.: Numerical analysis of blood flow in the heart.
\newblock J. Comput. Phys. \textbf{25}, 220--252 (1977)

\bibitem{Peskin:2002}
Peskin, C.S.: The immersed boundary method.
\newblock Acta Numerica \textbf{11}, 479--517 (2002)

\bibitem{Press:1992}
Press, W.H., Flannery, B.P., Teukolsky, S.A., Vetterling, W.T.: Fast fourier
  transform.
\newblock Ch. 12 in Numerical Recipes in FORTRAN: The Art of Scientific
  Computing \textbf{2}, 490--529 (1992)

\bibitem{Randall:1980}
Randall, D.J., Davie, P.S.: The hearts of urochordates and cephalochordates.
\newblock Comparative Anatomy and Development \textbf{1}, 41--59 (1980)

\bibitem{Reckova:2003}
Reckova, M., Rosengarten, C., deAlmeida, A., Stanley, C.P., Wessels, A.,
  Gourdie, R.G., Thompson, R.P., Sedmera, D.: Hemodynamics is a key epigenetic
  factor in development of the cardiac conduction system.
\newblock Circ. Res. \textbf{93}, 77 (2003)

\bibitem{SJones:2015}
S.~K~Jones R.~Laurenza, T.L.H.B.E.G.L.A.M.: Lift- vs. drag-based for vertical
  force production in the smallest flying insects.
\newblock J. Theor. Biol. \textbf{384}, 105--120 (2015)

\bibitem{Mahur:2012}
S.~R.~Mathura L.~Sunb, S.D.J.Y.M.: Application of the immersed boundary method
  to fluid, structure, and electrostatics interaction in mems.
\newblock Numerical Heat Transfer, Part B: Fundamentals: An International
  Journal of Computation and Methodology \textbf{62}, 399--418 (2012)

\bibitem{Samson:2007}
Samson, O.: A review of valveless pumping: History, applications, and recent
  developments (2007).
\newblock
  \urlprefix\url{http://www.researchgate.net/publication/267300626_A_Review_of_Valveless_Pumping_History_Applications_and_Recent_Developments}

\bibitem{Santhanakrishnan:2011}
Santhanakrishnan, A., Miller, L.A.: Fluid dynamics of heart development.
\newblock Cell Biochem. Biophys. \textbf{61}, 1--22 (2011)

\bibitem{Tucker:1988}
Tucker, D.C., Snider, C., Jr, W.T.W.: Pacemaker development in embryonic rat
  heart cultured \emph{in oculo}.
\newblock Pediatric Research \textbf{23}, 637--642 (1988)

\bibitem{Tytell:2010}
Tytell, E., Hsu, C., Williams, T., Cohen, A., Fauci, L.: Interactions between
  internal forces, body stiffness, and fluid environment in a neuromechanical
  model of lamprey swimming.
\newblock Proc. Natl. Acad. Sci. \textbf{107}, 19,832–19,837 (2010)

\bibitem{Waldrop:15BMMB}
Waldrop, L.D., Miller, L.A.: Large-amplitude, short-wave peristalsis and its
  implications for transport.
\newblock Biomechanics and Modeling in Mechanobiology pp. 1--14 (2015)

\bibitem{Weinbaum:2003}
Weinbaum, S., Zhang, X., Han, Y., Vink, H., Cowin, S.: Mechanotransduction and
  flow across the endothelial glycoalyx.
\newblock PNAS \textbf{100}, 7988--7995 (2003)

\bibitem{Zhu:2011}
Zhu, L., He, G., Wang, S., Miller, L.A., Zhang, X., You, Q., Fang, S.: An
  immersed boundary method by the lattice boltzmann approach in three
  dimensions.
\newblock Computers and Mathematics with Applications \textbf{61}, 3506–3518
  (2011)

\end{thebibliography}

\end{document}